\documentclass[aip,reprint]{revtex4-1}

\usepackage{dcolumn}
\usepackage{bm}

\usepackage[usenames,dvipsnames]{color}
\usepackage[toc,page]{appendix}
\usepackage{epsf}
\usepackage{bm}
\usepackage{dcolumn}
\usepackage{latexsym}
\usepackage{amsmath}
\usepackage{amsfonts}
\usepackage{amssymb}
\usepackage{graphicx}
\usepackage{braket}
\usepackage{multirow}
\graphicspath{}
\newcommand{\be}{\begin{eqnarray}}
\newcommand{\ee}{\end{eqnarray}}

\definecolor{darkred}{rgb}{.8,0,0}

\definecolor{darkblue}{rgb}{0,0,.7}

\usepackage[dvipsnames]{xcolor}

\usepackage[makeroom]{cancel}

\begin{document}


\title{Secure quantum remote sensing without entanglement} 



\author{Sean W. Moore}
\email{S.W.Moore@sussex.ac.uk}
\affiliation{Department of Physics and Astronomy, University of Sussex, Brighton, BN1 9QH, United Kingdom}

\author{Jacob A. Dunningham}
\email{J.Dunningham@sussex.ac.uk}
\affiliation{Department of Physics and Astronomy, University of Sussex, Brighton, BN1 9QH, United Kingdom}


\date{\today}

\begin{abstract}

Quantum metrology and quantum communications are typically considered as distinct applications in the broader portfolio of quantum technologies. However, there are cases where we might want to combine the two and recent proposals have shown how this might be achieved in entanglement-based systems. Here we present an entanglement-free alternative that has advantages in terms of simplicity and practicality, requiring only individual qubits to be transmitted. We demonstrate the performance of the scheme in both the low and high data limits, showing quantum advantages both in terms of measurement precision and security against a range of possible attacks.

\end{abstract}


\maketitle 

\section{Introduction}

Two of the most promising quantum technologies are quantum metrology and quantum communications. In the former, quantum correlations are used to measure quantities with a precision beyond what could be achieved by any classical means with the same resources \cite{Giovannetti2011-ma}; in the latter, the properties of quantum states are used to create secure communication channels \cite{Pirandola:20}. There are situations where we may want to combine both of these technologies in a single scheme enabling us to make precise measurements of remote quantities in a way that the information is completely secure from anyone who intercepts the signal or even has access to the remote sensor. 

Entanglement has been used to develop theoretical models to perform various remote sensing tasks without giving information away to a third party. Examples include detecting the relative positioning of two parties\cite{Giovanetti2002a} or non-zero magnetic fields in a network of detectors\cite{Kasai2022a}, the creation of a secure network of atomic clocks\cite{Komar2014a}, phase estimation strategies using solid state detectors\cite{Takeuchi2019a,Okane2020a}, a general qubit quantum resource with a trusted third party\cite{Huang2019a} and high dimensional probes\cite{Xie2018a}. 

The estimation of a parameter held at a remote site by a party who can be trusted to follow instructions, but who we do not want to know the measurement outcome, has been performed experimentally\cite{Yin2020a} using entangled photons. We propose a protocol that functions under similar conditions but does not require entanglement, using instead general qubits to allow one party, Alice, to estimate a phase at the remote location of the other, Bob. Possible issues with this scheme are that it relies on Bob being trusted to follow the protocol and the fact that Bob could always make his own independent measurement of the phase, hence circumventing the secrecy of the phase. However, there are scenarios where this is still useful. One example is a doctor, Alice, monitoring the health of a patient, Bob. It is in Bob's interests to follow the instructions so that the scheme works. However, if he does decide to make his own independent measurement, there is no issue with Bob having access to his own medical data. The key thing is that no third party can access it and that is what this scheme ensures. If Bob avoids making his own measurement, then he also ensures that his device is safe from being hacked.

We present a secure quantum remote sensing scheme (SQRS) that can estimate a phase and does not require entanglement, with the practical advantage that single qubit states may be easier to prepare. However, an equivalent scheme could be made with entanglement by having Alice and Bob share Bell pairs with Alice then measuring her part to drive the state of Bob's qubits. Similarly to existing SQRS schemes~\cite{Takeuchi2019a,Yin2020a}, we quantify the security by showing an asymmetry in Fisher information \cite{Fisher,Fihser_multiparticle_entanglement} of Alice and an eavesdropper, Eve, who can access Bob's measurement results on the parameter of interest, $\phi$, because they are sent through a public classical communication channel. Eve cannot gain any useful information about $\phi$, when her relevant Fisher information is zero. Then, similarly to other existing SQRS schemes~\cite{Huang2019a,Xie2018a} we show further security for attacks in the quantum communication channel by showing that in a noiseless implementation of the scheme it is statistically unlikely for Eve to interact with qubits in flight between Alice and Bob without revealing her presence. Finally, we introduce how shared secrets may be used to stop man in the middle attacks and how our scheme can be protected against some attacks that involve changing the classical data travelling from Bob to Alice.

\begin{figure*}
    \includegraphics[scale=1]{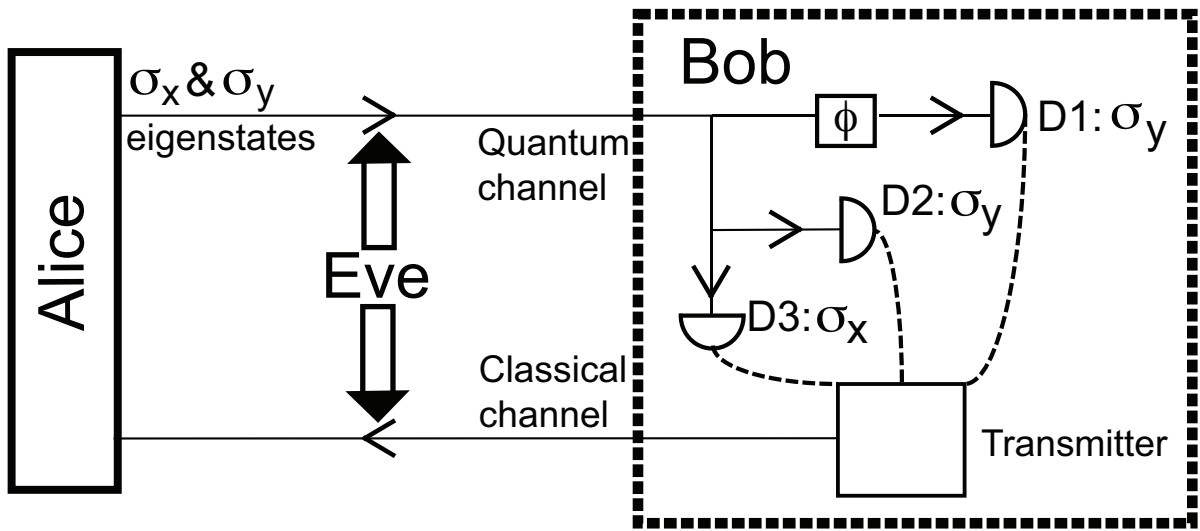}
    \caption{Scheme of the secure remote quantum sensing protocol. Alice sends eigenstates of $\sigma_x$ and $\sigma_y$ randomly through a quantum channel to Bob. Bob either encodes the parameter of interest, $\phi$, on the qubit and measures at D1 in the $\sigma_y$ basis or measures in the $\sigma_y$ or $\sigma_x$ basis D2 and D3 without encoding $\phi$. The measurement results and the detectors they correspond to are sent to Alice through a classical channel. Eve can attack the quantum and classical channel to try to gain information about $\phi$.}
    \label{fig:scheme}
\end{figure*}

We begin by presenting the basics of our protocol in Section~\ref{protocol}. We use quantum and classical Fisher information to find the states and measurements that optimise the sensitivity of the phase estimation. We then consider the security of the scheme starting with the classical channel. We use the classical Fisher information to show that Eve can learn nothing about $\phi$ from Bob's measurement results and the quantum Fisher information to show that she would learn nothing no matter the measurements Bob makes. We then show how it is statistically unlikely for Eve to evade detection if she tries to measure more than a few qubits travelling between Alice and Bob.

Next we consider resource efficiency, firstly analysing the effectiveness of the protocol's parameter estimation using limited resources by showing the variation of width of the likelihood function and the bias of an estimator for limited data. We then show how multiple passes through a sample can be used to approach the Heisenberg scaling without using entanglement.

In Section~\ref{photon_splitting} we consider a practical security issue that arises when using photons as the quantum resource, the problem of photon-splitting attacks. This vulnerability arises if weak coherent states are used as a more practical alternative to ideal single photon sources in schemes such as BB84 \cite{Bennett2014QuantumCP} since they could allow an eavesdropper to split off and measure copies of some states as they travel between Alice and Bob, without being detected. To mitigate this, photon-based schemes often use strategies such as decoy states \cite{Decoy_state_origin,Decoy_state_practical,Decoy_state_first_exp} as a way of improving security. We show that our protocol is significantly more robust to photon-splitting attacks than BB84 and may not require the operational overhead of these additional security measures.

In Section~\ref{security} we discuss further security features for a range of attacks, beginning with how shared secrets can be naturally applied to this scheme to ensure that Eve cannot imitate Alice or Bob in a man in the middle attack. We then introduce precautions to better protect against attacks where Eve may change some of the classical information sent from Bob to Alice. We consider the detection of spoofing attacks and show that, in the limited data regime of interest, our scheme maintains a high degree of security when Eve tries to hide her attacks on the quantum communication channel by manipulating the classical information Bob sends Alice.

\section{Protocol}\label{protocol}

The basic SQRS scheme is illustrated in Fig.~\ref{fig:scheme}. We consider a situation where Alice wants to make a measurement remotely at Bob's location without revealing the result to Bob or any eavesdropper, Eve. Alice and Bob share both a public classical communication channel and a quantum communication channel, each of which may be subject to eavesdropping or other external influences. As with previous schemes \cite{Takeuchi2019a, Okane2020a, Yin2020a}, we assume that Bob can be trusted to follow Alice's instructions. However, he is not required to be a secure node so, his classical information may be stolen without repercussions. All details of the scheme can be known publicly apart from what state Alice chooses on any given realisation and, of course, the value of the parameter $\phi$ being measured. In section~\ref{photon_splitting} we consider states encoded in photons but, the scheme is general and could be applied to other qubits. 

Alice sends appropriately chosen quantum states to Bob through an insecure quantum channel. Here, we take these to be the eigenstates of the Pauli $\sigma_x$ and $\sigma_y$ operators, for reasons discussed in \ref{Alice_info}. The range of different states and their probabilities can be public but the state of each particular instance is kept hidden by Alice. We assume that there is sufficient timing and authentication agreement for Alice and Bob to agree on which qubit is which.

At Bob's end the qubits are sent down one of two paths. The first of these paths encodes the parameter of interest, $\phi$, on the quantum state and then a measurement is made at detector D1 in the $\sigma_y$ basis.  The second path measures the state directly in the  $\sigma_y$  and $\sigma_x$ basis at D2 and D3 without encoding $\phi$. This serves as a test to verify the fidelity of the qubit states. The outcomes for each measurement and the corresponding detector are sent to Alice publicly through the classical communication channel. Alice performs a Bayesian analysis of the results of each path both to check the fidelity of the states arriving at Bob and to estimate the unknown parameter.

The detector that each qubit travels to is chosen at random once the qubit arrives at Bob and Eve can no longer interact with it. This ensures that Eve cannot selectively interact with qubits that will be used for parameter estimation. When interacting with qubits in the quantum channel she must take the risk that the qubits that she interacts with are sent to fidelity checking detectors and her presence would be revealed.

\subsection{Alice's information} \label{Alice_info}

In order to understand how much information Alice gains in the SQRS scheme, let us consider a general pure state qubit $\cos(\alpha/2)\ket{0} + \sin(\alpha/2)e^{i\beta}\ket{1}$ with $\alpha,\beta \in\Re$, that she sends to Bob. This is passed through a phase gate at Bob's end with the unknown parameter $\phi$, changing the state to $\cos(\alpha/2)\ket{0} + \sin(\alpha/2)e^{i\delta}\ket{1}$, where $\delta=\beta+\phi$. Since Alice knows $\alpha$ and $\beta$, any publicly transmitted measurement data about $\delta$ would enable Alice to find $\phi$, but any party without knowledge of $\beta$ would not be able to do so.

Let us now suppose that Bob measures the photon at D1 in the basis $\{+1,-1\} =\{\cos(\gamma/2)\ket{0} + \sin(\gamma/2)e^{i\varepsilon}\ket{1}, \sin(\gamma/2)\ket{0} - \cos(\gamma/2)e^{i\varepsilon}\ket{1}\}$. The probabilities of these two outcomes are
\begin{equation}
    P(\pm1|\phi) = \frac{1}{2}\left(1\pm\cos(\alpha)\cos(\gamma)\pm\sin(\alpha)\sin(\gamma)\cos(\zeta)\right), \label{meas_prob}
\end{equation}
where $\zeta = \beta + \phi -\varepsilon$. 

In the asymptotic limit of many measurements, $\mu$, the precision with which Alice can estimate $\phi$ is given by the classical Fisher information 
\begin{equation} \label{class_fisher}
    \mathcal{I}(\phi) = \sum_{i}\frac{1}{P(i|\phi)}\left(\frac{\partial P(i|\phi)}{\partial\phi}\right)^2,
\end{equation}
and the corresponding Cram{\'e}r-Rao bound
\begin{equation}
    \delta\phi\geq\frac{1}{\sqrt{\mu \mathcal{I}(\phi)}}.
\end{equation}
The classical Fisher information for the measurement Bob makes is 
\begin{widetext}
\begin{equation}
    \mathcal{I}(\phi) = \frac{\sin^2(\alpha)\sin^2(\gamma)\sin^2(\zeta)}{1-\cos^2(\alpha)\cos^2(\gamma)-\sin^2(\alpha)\sin^2(\gamma)\cos^2(\zeta) - \cos(\alpha)\cos(\gamma)\sin(\alpha)\sin(\gamma)\cos(\zeta)}.
\end{equation}
\end{widetext}
This has its optimal value of unity when $\alpha = \gamma = \pi/2$, corresponding to states and measurements in the $\sigma_x$-$\sigma_y$ plane of the Bloch sphere. The quantum Fisher information for the general pure state we are considering, $\cos(\alpha/2)\ket{0} + \sin(\alpha/2)e^{i\beta}\ket{1}$, is also unity, meaning that this scheme is optimal over all possible measurements because the classical Fisher information saturates the quantum Fisher information. Alice and Bob therefore choose to operate in this plane. The same choice of states and measurements is effective on the test path for verifying that the fidelity of the states is maintained through the quantum channel.

\begin{figure}[ht]
    \centering
    \includegraphics[width=.45\textwidth]{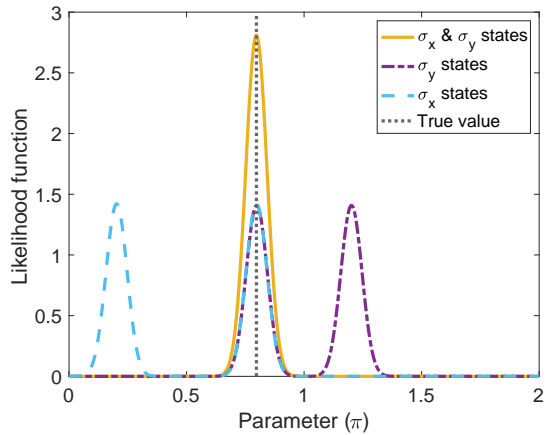}
    \caption{Alice's mean likelihood function as given by Eq.~\ref{Prob_eq} with $\mu=100$ (averaged over $10^3$ realisations) and compared with the true value of $\phi$. Results are shown when Alice only sends $\sigma_x$ states or $\sigma_y$ states. In both these cases there is a periodicity over the $2\pi$ range, which creates an identifiability problem. By using both $\sigma_x$ and $\sigma_y$ states a single peak corresponding to the true value can be identifed.}
    \label{fig:identifiability_01}
\end{figure}

To maintain security, Alice could use any set of symmetric states in this plane and corresponding probabilities so that they average to an identity density matrix.  For consistency with previous work, we take Alice to use eigenstates of the $\sigma_x$ and $\sigma_y$ operators with equal probabilities, which she sends to Bob who makes measurements in the $\sigma_y$ basis. The probabilities for Bob to obtain the two different measurement outcomes, given the four possible states that Alice sends, are given in table~\ref{table:probs}
\begin{table}
\centering
\caption{The measurement probabilities $p_j$ when Alice sends $\sigma_x$ and $\sigma_y$ eigenstates to Bob who encodes $\phi$ on them and measures in the $\sigma_y$ basis. The number of times that Bob gets a result corresponding to each $p_j$ is $n_j$.}
\begin{tabular}{ |c|c|c|  } 
 \hline
 \multirow{2}{6em}{\bf Eigenstate} &  \multicolumn{2}{c|}{\bf Measurement outcome}  \\
& $\sigma_y = +1$ & $\sigma_y = -1$  \\
 \hline
$ \sigma_x = +1$  &$p_1 = \frac{1}{2}(1+\sin\phi)$   &$ p_2 = \frac{1}{2}(1-\sin\phi)$\\[3pt]
$ \sigma_x = -1$  &   $p_3 =  \frac{1}{2}(1-\sin\phi)$  & $p_4 = \frac{1}{2}(1+\sin\phi)$ \\[3pt]
$ \sigma_y = +1$  & $p_5= \frac{1}{2}(1+\cos\phi)$ & $p_6= \frac{1}{2}(1-\cos\phi)$ \\[3pt]
$ \sigma_y = -1$   & $p_7= \frac{1}{2}(1-\cos\phi)$ & $p_8= \frac{1}{2}(1+\cos\phi)$ \\[3pt]
 \hline
\end{tabular}\label{table:probs}
\end{table}
\raggedbottom
Suppose that in a given experiment with $\mu$ measurements, the number of results for each of the outcomes with probabilities given in table~\ref{table:probs} is $\{n_j\}$, where $j\in\{1,2,...8\}$ and $\mu = \sum_j n_j$, a Bayesian approach gives us Alice's likelihood function for $\phi$,
\begin{eqnarray}
    \mathcal{L}(\phi) \propto&& (1+\sin\phi)^{n_1+n_4}(1-\sin\phi)^{n_2+n_3}\nonumber \\
    &&\times(1+\cos\phi)^{n_5+n_8}(1-\cos\phi)^{n_6+n_7}.
    \label{Prob_eq}
\end{eqnarray}
It can been seen from table~\ref{table:probs} that the measurement outcome probabilities for the $\sigma_x$ and $\sigma_y$ eigenstates depend only on $\sin(\phi)$ and $\cos(\phi)$ respectively. This means that any estimation based solely on one basis such as those used previously\cite{Takeuchi2019a} for the detection of small phases, can only estimate the parameter in a $\pi$ range because they give rise to symmetric likelihoods in the $2\pi$ range as shown in Fig.~\ref{fig:identifiability_01}. This ambiguity is removed by using two sets of states that are perpendicular to each other in the Bloch sphere, allowing for estimation over the full $2\pi$ range. An additional reason for using both $\sigma_x$ and $\sigma_y$ states is security. If Alice always sends states from just one of these sets e.g. $\sigma_x=\pm1$, and that set is publicly known, Eve could measure in this basis without changing the state and so implement a measure and resend attack without being detected.

\subsection{Eve's information}\label{Eve_info}

Security of the classical information is maintained because Alice is the only party to know the state of each qubit she sends. If Alice sends each of the $\sigma_x$ and $\sigma_y$ eigenstates with equal probability she can ensure that Eve gains no information from the measurement outcomes that Bob sends through the classical channel. We can see this from the probabilities in table~\ref{table:probs}. Since Eve does not know the state, the probability of a $+1$ or $-1$ measurement outcome is given by taking an equally weighted sum of the probabilities in the corresponding column of table~\ref{table:probs}, i.e 
\begin{equation}
    P_{Eve}(+1|\phi) = P_{Eve}(-1|\phi)  =\frac{1}{2}.
\end{equation}
Since these are independent of $\phi$, their derivative with respect to $\phi$ vanishes and Eve's classical Fisher information, as given by Eq.~(\ref{class_fisher}), is zero. Similarly, Eve's density matrix for each photon is $\hat{\rho}_{\text{Eve}} = \frac{1}{2} I$ both before and after interaction with $\phi$, giving a quantum Fisher information of zero. This means that Eve can gain no information from the classical channel regardless of the measurements that Bob performs. Comparing this with \ref{Alice_info} where Alice obtains the maximum possible information about $\phi$, we see a clear information asymmetry between Alice and Eve that drives this scheme.

Eve cannot learn anything about the parameter $\phi$ from the classical information that Bob has and sends through a public channel because she does not know what any of the qubit states are when they arrive at Bob. However, if she were to interact with the qubits in some way as they travel from Alice to Bob, she may be able to determine the state of some of them. For instance, she could block some qubits and replace them with her own or she could perform a measure and resend attack as qubits pass from Alice to Bob.

Due to the indistinguishability of the quantum states travelling between Alice and Bob, Eve cannot interact with them without risking changing them. If Bob chooses which qubits are used for parameter estimation or state fidelity checking at random once Eve can no longer interact with them he ensures that Eve cannot selectively target only those states that she knows will not be used to test for her presence. As suggested in another SQRS scheme~\cite{Huang2019a}, by having a similar set of qubits travel through a quantum channel as in the BB84 quantum cryptography protocol~\cite{Bennett2014QuantumCP} we may retain similar security.

We demonstrate here that it is statistically unlikely for Eve to attack the quantum channel without being detected by Alice. Therefore, a noiseless implementation this scheme is secure against such attacks. For example, if Eve were to perform a maximal discrimination measurement on a qubit in flight from Alice to Bob and replace it with her measurement result she would have a probability of 1/2 of successfully determining what state Alice had sent~\cite{Zhang2008a}. This is also a maximal disturbance measurement so Eve has a probability 1/2 of sending a state in the wrong eigenbasis to Bob. If this qubit travels to the detector D2 or D3 that corresponds to the original eigenbasis that Alice sent the qubit in, there is a probability of 1/2 of giving a result that does not correspond to that qubit and signals the possibility of such an attack. Thus, each qubit that Eve measures in this way while in transit and is replaced by another that travels to the detector D2 or D3 corresponding to the eigenbasis of the original state has a probability of $P_{\text{sing}}=1/4$ of signalling the attack to Alice. Over many attempts the probability of disagreement, $P_{\rm \text{dis}}$, between an original state and Bob's fidelity checking measurement is
\begin{equation}
P_{\rm \text{dis}} = 1 - \left(1-P_{\rm \text{sing}})\right)^{N},
\end{equation}
where $P_{\rm \text{sing}}$ is the probability of a disagreement for any attack on a single state and $N$ is the total number of fidelity checking measurements. We take the rate at which Bob sends qubits down the fidelity checking path to D2 and D3 as $0<p<1$. With each fidelity checking detector equally likely, the probability of Bob sending a qubit to each detector D2 or D3 is $p/2$ and the parameter estimation detector at D1 is $1-p$. When Eve attacks $\mu$ qubits the expected number that travel to the corresponding detector is $\mu p/2$ for all the initial states. Thus, the expected disagreement is 
\begin{equation}
    P=1-(3/4)^{\mu p/2}. \label{eq:disagree}
\end{equation}
In our protocol, we take the simple approach that as soon as Alice detects a discrepancy she stops the scheme because she cannot be sure that there is no eavesdropper. We see from Eq.~(\ref{eq:disagree}) that the probability of Eve being detected by this method exponentially approaches unity with the number of states attacked. Any attack that changes the states of the qubits in the quantum channel, such as spoofing Alice's results by adding a phase, would have a similar exponential detection rate.

\subsection{Limited data}

When there are a large number of measurements such as in \cite{Yin2020a}, the Fisher information and Cram\'er-Rao bound can be used to quantify the precision with which Alice can determine $\phi$. However, we may want to explore what happens when we are not in that limit, not least because sending many copies of the same information represents a security risk; Eve would only need to be able to access a tiny fraction of the copies to still be able to gain significant information about $\phi$.
Previous studies have explored how quantum metrology can be applied when there are a limited number of measurements \cite{Rubio_2018,Rubio_2019,Rubio_2020} and the remainder of this paper will focus on that regime. 

For a set of measurement results, $\{n_i\}_{i=1}^{8}$, the likelihood function is given by Eq.~\ref{Prob_eq}. In the asymptotic regime, this becomes an increasingly narrow normal distribution with a mean corresponding to the true value of $\phi$ and a variance given by the inverse classical Fisher information. This makes the maximum likelihood function a useful estimator. However, with limited data the likelihood function may be biased and have a large variance, meaning we need to account for the fact that it has a circular support with a period of $2\pi$. 

When a likelihood function is non-negligible over the $2\pi$ support, the choice of where to cut the circular support to superimpose it on a flat support and the wrapping of the likelihood function become important and affect the statistics drawn from that distribution~\cite{DirectionalStats}. Previous theoretical work was restricted to large data and parameters close to zero~\cite{Takeuchi2019a} so did not need to take this into account. To best analyse our limited data likelihood functions we use the circular analogue of the standard deviation $\nu$ which approximates the linear standard deviation for narrow distributions, as discussed in Appendix \ref{circStats}.

\begin{figure}[ht]
    \centering
    \includegraphics[width=.45\textwidth]{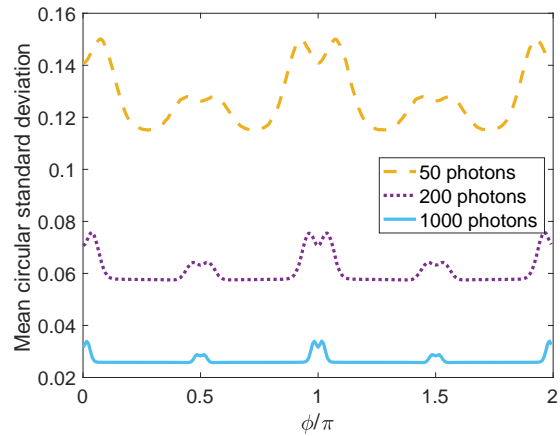}
    \caption{The mean circular standard deviation of Alice's likelihood function as a function of the true value of $\phi$, shown for different total numbers of qubits, $\mu$, used on the measurement path, D1. This shows that the width of the distribution depends on $\phi$, but this dependence reduces as the number of measurements increases.}
    \label{fig:circDist}
\end{figure}

\begin{figure}[ht]
    \centering
    \includegraphics[width=.45\textwidth]{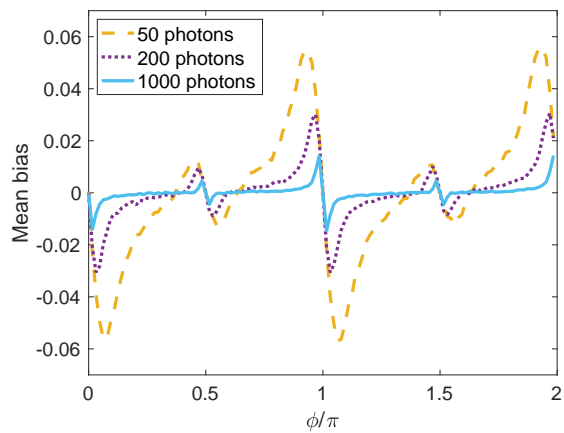}
    \caption{The mean bias of Alice's maximum likelihood estimator from the true value of $\phi$, shown as a function of $\phi$ for different total numbers of qubits, $\mu$, used on the measurement path, D1. The bias depends on $\phi$ but decreases as the number of measurements increases.}
    \label{fig:meanBias}
\end{figure}

We begin by considering a measure of the width of the distribution. For limited data, the circular standard deviation varies with the true value of the parameter $\phi$ as shown in Fig.~\ref{fig:circDist}. As expected, the dependence on $\phi$ reduces as the number of measurements increases which is consistent with the classical Fisher information being independent of the value of the parameter.

Next we consider how having limited data affects the bias of Alice's likelihood functions. Fig.~\ref{fig:meanBias} shows the relationship between the bias of the maximum likelihood estimator and the true value of $\phi$. A comparison of Fig.~\ref{fig:circDist} and Fig.~\ref{fig:meanBias} shows that, for all values of $\phi$, the bias is always much smaller than the width of the likelihood function. This means that Alice's Bayesian method of estimating $\phi$ should also work well in the low data regime.

Alice can optimise her estimation by operating in regions of Figs.~\ref{fig:circDist} and \ref{fig:meanBias} that minimise both the width of the likelihood function and the bias, i.e. avoiding regions where $\phi$ is close to half-integer multiples of $\pi$. She can do this if she has some prior information about $\phi$. Alternatively, as the measurement progresses, Alice will build up knowledge of $\phi$ and can use this to shift to a preferred operating region. In the $\sigma_x$-$\sigma_y$ plane the probabilities of the results can be found by substituting $\alpha = \gamma =\pi/2$ into Eq.~\ref{meas_prob} to give
\begin{equation}\label{eq:variable_probability}
    P(\pm1) = \frac{1}{2}\left( 1 \pm \cos(\beta+\phi-\varepsilon)\right).
\end{equation}
If Alice wants to shift her peak by $\eta$ she can rotate all of her initial states $\beta\in \{0,\pi/2,\pi,3\pi/2\}$ to $\beta\in \{-\eta,\pi/2 - \eta,\pi - \eta,3\pi/2 - \eta\}$. This would be undetectable to all other parties and would not reveal any information she has about $\phi$.

\subsection{Quantum enhancement}

So far we have focused on how the quantum features of the scheme can be used for security, however they can also be used to give an enhancement in the measurement precision itself. This is usually achieved by making use of entangled states to improve how the uncertainty in the parameter, $\Delta\phi$, scales with $N$, the number of particles used\cite{Giovannetti2011-ma}. For unentangled particles, this goes as the standard quantum limit $\Delta\phi\sim 1/\sqrt{N}$, but with entanglement it is possible to achieve a Heisenberg scaling $\Delta\phi\sim 1/{N}$. A standard approach is to use NOON states that were developed~\cite{Boto2020a}, popularised~\cite{Dowling2008a}, and named~\cite{Lee2002a} by Jonathan Dowling. These have the form
\begin{equation}
    |\psi\rangle = \frac{1}{\sqrt2}\left( \ket{N,0} + \ket{0,N}\right).
\end{equation}
If one mode is subjected to a phase, $\phi$, this adds coherently for all $N$ particles, giving
\begin{equation}
    |\psi\rangle = \frac{1}{\sqrt2}\left( \ket{N,0} + e^{iN\phi}\ket{0,N}\right),
\end{equation}
which has a quantum Fisher information with respect to $\phi$ of $N^2$, leading to a possible scaling in the measurement precision of $\Delta\phi\sim 1/{N}$.
NOON states have the disadvantage of being fragile to loss and difficult to create. Despite this, experiments have demonstrated them in the laboratory and shown their improved scaling for measurements~\cite{Mitchell2004a,Matthews2011a,Leibfried2005a}. However, it is not clear how they can be applied to our SQRS scheme. Instead we consider an entanglement-free scheme for Heisenberg-limited phase estimation that is very much in the spirit of our SQRS protocol and can easily be applied to it~\cite{Higgins2007a}.

The idea behind this quantum enhancement is simply that Bob passes the state he receives from Alice through $\phi$ multiple times 
or interacts the qubit with the phase containing Hamiltonian for a longer time before measuring it. For $m$ passes (or an $m$-fold increase in interaction time) this replaces $\phi$ with $m\phi$ in the states and corresponding detection probabilities, meaning that the Fisher information is enhanced by a factor of $m^2$, as can be seen from Eq.~\ref{class_fisher}. This is a multiplicative factor so it would not allow Eve to gain any knowledge of the parameter $\phi$ under the same conditions that give her a Fisher information of zero for single passes through the sample.

\begin{figure}[ht]
    \centering
    \includegraphics[width=.45\textwidth]{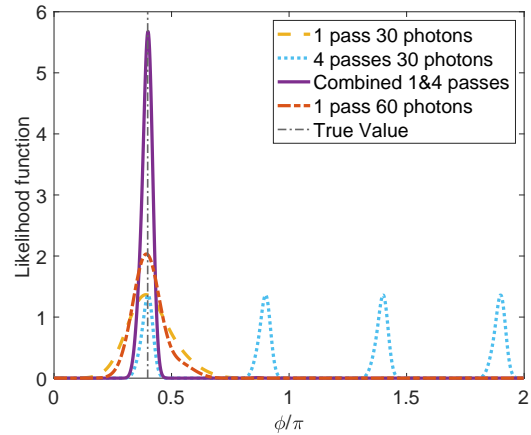}
    \caption{The mean likelihood functions averaged averaged $10^{4}$ times for different single-pass and $4$-pass measurement strategies. Combining the results from 30 qubits with a single pass and 30 qubits with 4 passes gives a significant advantage over using all 60 qubits for a single pass test.}
    \label{fig:mean_multipass_example}
\end{figure}

\begin{figure}
    \centering
    \includegraphics[width=.45\textwidth]{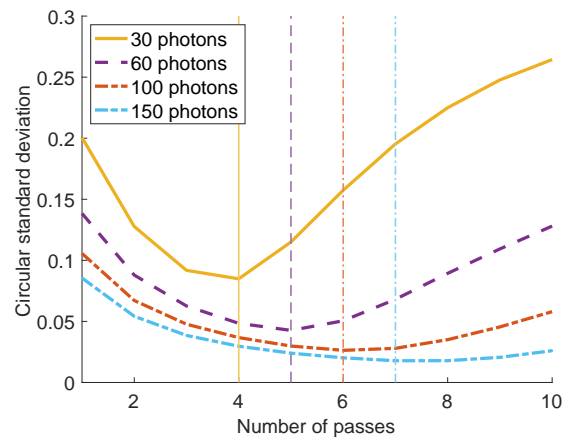}
    \caption{The mean circular standard deviation when averaged over many possible true values for the mean likelihood function when combining the results of a single pass and a multipass test each with the given number of qubits. Vertical lines show the minima. The widths initially decrease as the number of passes increases due to the increasing narrowness of the likelihood functions with more passes. However, they then increase again once the number of passes is too great and the single pass test begins picking out more than one multipass test peak. The vertical lines mark where the minimum circular standard deviation is achieved.}
    \label{fig:multipass_optimal_limited_data}
\end{figure}

While this is a simple and convenient way of gaining an $m$-fold enhancement in measurement precision, it has the disadvantage of creating $m$ equally spaced peaks in the likelihood function over a $2\pi$ range. The problem with this is that Alice can be left with an identifiability problem where she knows that one of the peaks is correct but not which one. There are different ways of dealing with this~\cite{Berry2009a}. If Alice has prior information of width $2\pi/m$ she can simply ask Bob to perform $m$ passes giving her a quantum enhancement of $m^2$, while ensuring there is only a single peak. Another approach is to make measurements with different numbers of passes. A single peak can be achieved by combining information from measurements with numbers of passes that have no common factors and sufficient qubits that their peaks do not overlap. 

In Fig.~\ref{fig:mean_multipass_example} we illustrate this effect for the combination of single pass and a $4$-pass measurement. A single pass measurement with 30 qubits gives a relatively broad likelihood function around the true value of $\phi$; a $4$-pass measurement with 30 qubits gives narrower peaks, but with a four-fold multiplicity meaning that the true value cannot be identified. Combining a single pass and a $4$-pass with 30 qubits each (60 in total) gives the best of both worlds with a narrow peak at the right value of $\phi$. This is compared with the result using all 60 qubits in single pass measurements, which has a broader peak.

Fig.~\ref{fig:multipass_optimal_limited_data} illustrates the interplay between the number of qubits used and the minimum circular standard deviation that can be achieved for single-pass and $m$-pass strategies that use the same number of qubits for each test. Initially the standard deviation decreases as the number of passes increases. However, the standard deviation increases again if there are too many passes for the number of qubits. This is because the single-pass likelihood distribution will not always be narrow and sufficiently well-placed to pick out only the correct single peak from the $m$-pass distribution, the peaks of which are separated by $2\pi/m$. This leads to a distinguishability problem and a larger standard deviation.

By combining results from different numbers of passes Alice can get an $m^2$ enhancement to her information with a cost of using enough qubits on a single pass test to be sure that she picks out the correct peak.

\section{Photon-number splitting attacks}\label{photon_splitting}

We saw in \ref{Eve_info} that Eve gains no information about $\phi$ from the information communicated in the classical channel alone and if she tries to measure states in the quantum channel, she will be detected exponentially quickly. In the remainder of this paper, we will consider further security details of the scheme such as spoofing and `man in the middle' type attacks \cite{Fei2018a}. In this section, we focus on the viability of photons for the practical implementation of this scheme by considering photon-number splitting attacks\cite{Huttner1995a, Liu2011a} by an eavesdropper.

The security of the quantum channel so far has been based on the assumption that Alice sends ideal qubits such as perfect single-photon states. If a state sometimes contains more than one photon, there is the possibility of Eve skimming off a photon, without Alice or Bob knowing, and using it to gain information about $\phi$. There has, therefore, been a lot of interest in creating single-photon sources and current systems for realising this include colour centres, trapped atoms, quantum dots and heralded spontaneous parametric down conversion sources~\cite{Aharonovich2016}. While good progress is being made, none of these systems are ideal; they can be difficult to implement and suffer from some degree of multi-photon emissions and low flux rates.
Here we show that our protocol is significantly more secure to photon-splitting attacks that BB84 and so has less reliance on single-photon sources. 

In our scheme, we can consider highly attenuated weak coherent state sources. For practical schemes, it is important to consider the rate at which Alice gains information. By having a higher average photon number per state we increase the flux rate of photons arriving at Bob and therefore the bandwidth and information gain rate of Alice. However, this also increases the rate of there being two or more photons in a wave-packet and therefore the probability of Eve succeeding in a photon-number splitting attack. Current photon-based quantum key distribution schemes typically use decoy states to overcome this problem\cite{Decoy_state_origin,Decoy_state_practical,Decoy_state_first_exp}. We show here that our SQRS scheme leads to significant information asymmetry between Alice and Eve even in the presence of photon-number splitting attacks, especially when combined with singlepass-multipass combined estimation. 

The number distribution for photons in a coherent state is
\begin{equation}
    P_{coh}(k) = \frac{e^{-\bar{k}}\bar{k}^k}{k!},
\end{equation}
where $\bar{k}$ is the mean photon number. The rate at which Alice gains information is proportional rate at which there is at least one photon per state, i.e. $1-e^{-\bar{k}}$. Whenever there is more than one photon, there is the possibility of Eve gaining information. However, SQRS differs from BB84 in a key way. In BB84, Alice and Bob publicly reveal to each other what bases they used. If Eve manages to steal some photons, she is able to wait until this information is revealed and then measure her photons in the same basis to find some of the bits of the key. In SQRS, Alice and Bob never need to communicate any such information. So, even if Eve gains some photons without being detected, she does not know what basis to measure them in.

Here we compare the information gained by photon splitting in our scheme to that for the BB84 protocol. In BB84 where the measurement basis is revealed Eve can gain all the information about the split photons. The rate of Eve's information gain relative to Alice's in BB84 is therefore
\begin{equation} \label{relative_rate_BB84}
     \left(\frac{E}{A}\right)_{BB84} \leq \frac{\sum_{j=2}^{\infty} P_{coh}(j)}{\sum_{j=1}^{\infty} P_{coh}(j)}  = \frac{e^{\bar{k}} -1 - \bar{k} }{e^{\bar{k}} -1}.
\end{equation}
For our SQRS scheme, we consider that Eve has access to Bob's classical information as it is sent through a public communication channel. Therefore, if she has some information about the state of a photon that Alice sends to Bob she may use this and Bob's measurement result to gain some information about $\phi$.

Splitting photons enables Eve to gain a copy of the qubit being sent to Bob. By measuring this copy, Eve gains some information about Bob's state before its interaction with $\phi$. From this and Bob's publicly available measurement outcomes, Eve's Fisher information for $\phi$ can be non-zero.

To put limits on the relative information rate of Eve compared to Alice in SQRS we consider the quantum Fisher information when Eve is able to perfectly obtain any extra photons. If she splits off one photon she can perform any set of measurements on it. If she splits off more than one photon we make the simplifying assumption that she gains full information about the photon. This gives an upper bound to the relative information of Eve to Alice
\begin{eqnarray} \label{relative_rates}
     \left(\frac{E}{A}\right)_{SQRS} \leq \frac{\mathcal{F}_E(\phi) P_{coh}(2) + \sum_{j=3}^{\infty} P_{coh}(j)}{\sum_{j=1}^{\infty} P(j)} \nonumber \\ = 
      \frac{e^{\bar{k}} -1 - \bar{k} - \frac{1-\mathcal{F}_E}{2}\bar{k}^2}{e^{\bar{k}} -1},
\end{eqnarray}
where $\mathcal{F}_E(\phi)$ is Eve's quantum Fisher information for $\phi$ when she has a single copy of the initial state she has split off. For the four photon states that Alice can send $\{\ket{\sigma_x= \pm 1},\ket{\sigma_y=\pm 1}\}$ the probabilities for Eve to get the result, $+1$, corresponding to a projection onto $\cos(\delta/2)\ket{0} + \sin(\delta/2)e^{i\gamma}\ket{1}$, are
\begin{eqnarray}
    P(E_{+1}|\sigma_x=\pm 1) &&= \frac{1}{2} \left(1\pm \sin(\delta)\cos(\gamma)\right) \\
    P(E_{+1}|\sigma_y=\pm 1) &&= \frac{1}{2} \left(1\pm \sin(\delta)\sin(\gamma)\right).
\end{eqnarray}
Using Bayes' rule for Eve's posterior probability gives the probability she has a state $\ket{\Psi_j}$ given the measurement outcome $E_{+1}$
\begin{equation}
    P(\Psi_j | E_{+1}) = \frac{P(E_{+1} | \Psi_j) P(\Psi_j)}{\sum_j P(E_{+1} | \Psi_j) P(\Psi_j)},
\end{equation}
where $P(E_{+1}) = \sum_j P(E_{+1} | \Psi_j) P(\Psi_j) = \frac{1}{2}$ and $P(\Psi_j) = \frac{1}{4}$  for all $j$, since we take the probability of all of the initial states to be equal. Taking the specific case of the $\sigma_x$ and $\sigma_y$ eigenstates, we get
\begin{eqnarray}
    P(\sigma_x=\pm 1|E_{+1}) &&= \frac{1}{4} \left(1\pm \sin(\delta)\cos(\gamma)\right) \\
    P(\sigma_y=\pm 1|E_{+1})&&= \frac{1}{4} \left(1\pm \sin(\delta)\sin(\gamma)\right).
\end{eqnarray}

\begin{figure}[ht]
    \centering
    \includegraphics[width=.45\textwidth]{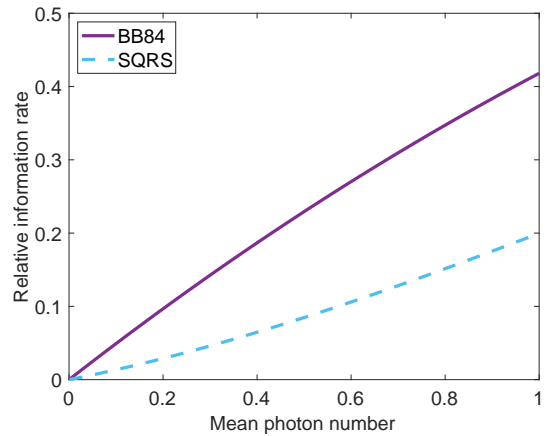}
    \caption{Upper bound to the information that Eve gains relative to Alice when making photon-number splitting attacks, as a function of the mean number of photons in each state. Results are compared for BB84 (solid line) and SQRS (dashed line). SQRS can be seen to be significantly more secure to these attacks than BB84.}
    \label{fig:photon_splitting_FI}
\end{figure}

Since each photon that Eve splits off is a copy of a photon that Bob measures, Eve can use her knowledge of the state from photon splitting to gain information about $\phi$. When Eve is able to measure a single copy of a photon's initial state her corresponding density matrix for that state is
\begin{eqnarray}
    \hat{\rho}_E = \frac{1}{2}\Big( I + \frac{1}{2} \sin(\delta)&&\cos(\phi+\gamma) \sigma_x \nonumber \\ &&+ \frac{1}{2} \sin(\delta)\sin(\phi+\gamma) \sigma_y\Big).
\end{eqnarray}
From this Eve has a quantum Fisher information, $\mathcal{F}_E$, for $\phi$ of
\begin{equation}
    \mathcal{F}_E = \frac{1}{4}\sin^2(\delta) \leq \frac{1}{4}.
\end{equation}
This has its maximum value when Eve measures in the same $\sigma_x$-$\sigma_y$ plane as Alice's states and is independent of the orientation, $\gamma$, of the projective measurement in that plane. Substituting $\mathcal{F}_E =1/4$ into Eq.~(\ref{relative_rates}) gives an upper bound to the information Eve gains from photon splitting attacks relative to Alice. These results are shown as a function of mean photon number per state in Fig.~\ref{fig:photon_splitting_FI} and compared with BB84. We see that SQRS is significantly less vulnerable to photon splitting attacks than BB84. The same results apply so long as Alice sends states with equal probability of having $\beta \in \{\beta_0,\beta_0+\pi/2,\beta_0+\pi,\beta_0+3\pi/2\}$, for any $\beta_0$, which is within her control.

If Alice uses a well-chosen multipass-singlepass combination such as those shown in Fig~\ref{fig:multipass_optimal_limited_data} the information asymmetry between Alice and Eve when performing photon splitting attacks will be further enhanced. Eve, with less information for the singlepass test than Alice will be unable to pick out only one peak or guarantee that she picks out the correct peak from the multipass test. Therefore, unless she has sufficient prior information to pick out the correct peak before starting, she will not be able to take advantage of the Heisenberg scaling that Alice gets using the multipass method.

When using limited data and a well-chosen multipass-singlepass combination Alice and Bob may decide to continue with the protocol regardless of the possibility of a man in the middle attack knowing that Eve would be able to extract a much smaller amount of information than Alice.

\section{Further security features}\label{security}

\begin{figure*}
    \includegraphics[width = 450pt]{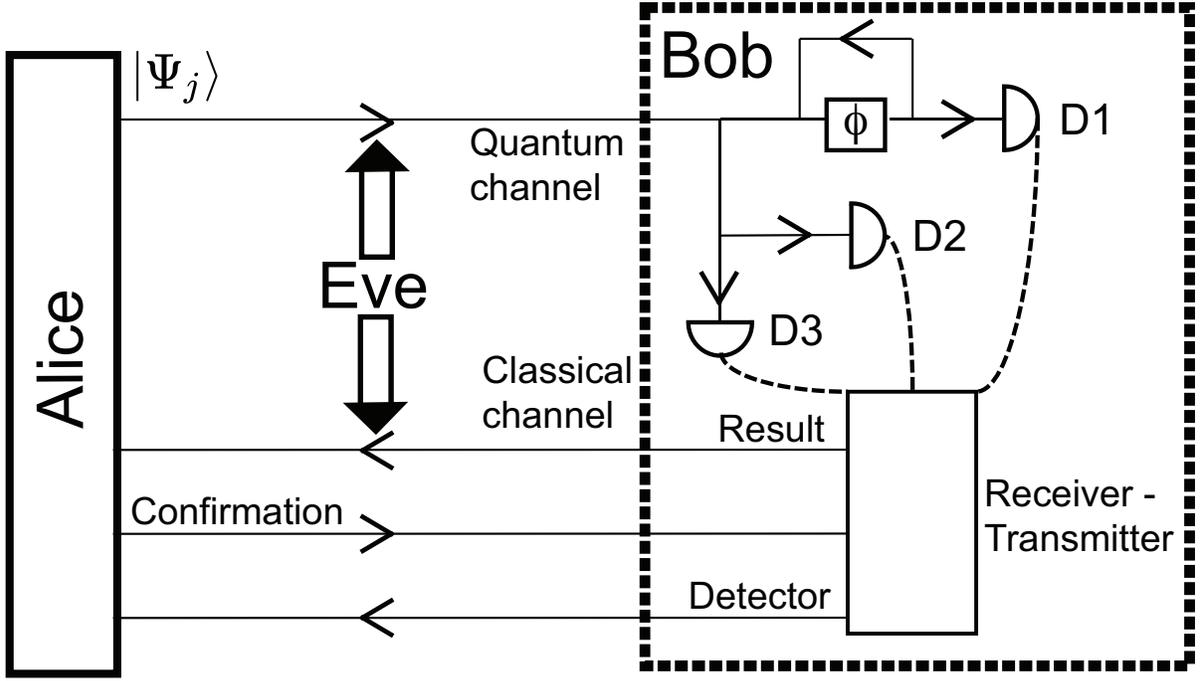}
    \caption{The SQRS scheme with additional security features. Alice sends states $\ket{\Psi_j} = \frac{1}{\sqrt{2}}\left(\ket{0} + e^{i\theta_j}\ket{1}\right)$, $\theta_j \in \{\varepsilon, \varepsilon+\pi/2,\varepsilon+\pi,\varepsilon+3\pi/2\}$ to Bob through the quantum channel, where $\varepsilon$ is a secret shared between Alice and Bob. On the test path Bob performs projective measurements onto $\frac{1}{\sqrt{2}}\left(\ket{0} \pm ie^{i\varepsilon}\ket{1}\right)$ at detector D2 and $\frac{1}{\sqrt{2}}\left(\ket{0} \pm e^{i\varepsilon}\ket{1}\right)$ at detector D3. On the measurement path Bob controls the number of times that the states pass through $\phi$ and then performs a projective measurement onto $\frac{1}{\sqrt{2}}\left(\ket{0}\pm  e^{i\tilde{\varepsilon}}\ket{1}\right)$ at detector D1, for some secret $\tilde{\varepsilon}$. Bob initially sends the measurement result to Alice through the classical channel keeping which detector that performed the measurement secret. On receipt of the measurement result Alice sends a confirmation message to Bob who then sends the information on which detector performed the measurement.}
    \label{fig:photon_SQRS}
\end{figure*}

In this section we show how some straightforward adaptations to our basic protocol, shown in Fig.~\ref{fig:scheme}, allow us to include a wider range of security features. The scheme with these additional security features is shown in Fig.~\ref{fig:photon_SQRS}. In Section~\ref{authentication} we discuss how shared secrets can be applied to the protocol to enhance its security. Sections \ref{spoofing} and \ref{MIM} provide a closer look at the classical communication channel with the former introducing delayed path information to protect against spoofing attacks and the latter putting a limit on man in the middle type attacks involving the classical communication channel.

\subsection{Shared secrets}\label{authentication}

In order for Alice and Bob to be sure they are communicating with the party they think they are, quantum key distribution requires channel authentication~\cite{Wang2021}. So far, we have assumed that Alice and Bob are able to authenticate their communication channels throughout this process. However, in case they fail and Eve attempts to imitate one to the other, shared secrets may be used to alert Alice to the deception and stop Eve from gaining any useful information about $\phi$ from Bob.

Suppose Alice and Bob share a secret value for an angle, $\varepsilon$, that Alice shifts her states by, i.e $\beta\in\{\varepsilon,\varepsilon+\frac{\pi}{2},\varepsilon+\pi,\varepsilon+\frac{3\pi}{2}\}$. If Bob then uses the angles $\varepsilon$ and $\varepsilon +\pi/2$ for his measurements on the test paths D2 and D3, the state checking relationships remain the same as if neither party had rotated by $\varepsilon$. This can be seen from Eq.~(\ref{eq:variable_probability}) where $\phi=0$ on the test paths and $(\beta -\varepsilon) \in\{0,\frac{\pi}{2},\pi,\frac{3\pi}{2}\}$. This means that Alice can tell if Eve tries to impersonate Bob because the test outcomes she sends Alice will not match what Alice expects. Eve will not be able to tell whether Alice has shifted her basis without measuring some of the states Alice sends and she would be quickly detected if she tried to do this. 

Similarly, Alice and Bob could share another secret $\tilde{\varepsilon}$ for the orientation of the detector D1 on the parameter measuring path. If Eve tries to impersonate Alice to Bob she will gain no meaningful information about $\phi$. Eve would control $\beta$ in this scenario and it can be seen from Eq.~(\ref{eq:variable_probability}) that she would gain information about $\phi-\tilde{\varepsilon}$. However, without knowledge of the secret value of $\tilde{\varepsilon}$ this tells her nothing about $\phi$. 

\subsection{Spoofing}\label{spoofing}

As shown in section~\ref{Eve_info}, any attempt to spoof Alice's results by adding a phase in the quantum channel would be detected by Alice. However, Eve could attempt to spoof the results and give Alice a false estimation of $\phi$ without her realising it by manipulating the $\{-1,+1\}$ data in the classical channel that correspond only to measurements of $\phi$, i.e. not test outcomes.
For example, if Eve swaps {\it every} such datum, Alice will be led to believe that the correct value of the parameter is $\phi + \pi$. This can be seen from Eq.~(\ref{Prob_eq}), where if we we make the swaps $\phi \to \phi +\pi$, $n_1 \leftrightarrow n_2$, $n_3 \leftrightarrow n_4$, $n_5 \leftrightarrow n_6$, $n_7 \leftrightarrow n_8$, the probability distribution is unchanged. 

By ensuring that Eve cannot know the path that was used or interact with Bob sending that information to Alice until after Alice has received the result, Bob guarantees that Eve cannot change the results for the measurement path without also changing the results of the test path which would reveal such an attack to Alice.

This is shown by the three classical information exchanges between Alice and Bob in Fig.~\ref{fig:photon_SQRS}. When a measurement is made Bob keeps which detector made the measurement a secret and sends only the measurement result $\{-1,+1\}$ to Alice. Once Alice has received the result she sends a message to Bob to confirm this. On receipt of this message, Bob sends the information of which detector and therefore, which path, the qubit was sent through. They proceed similarly for each qubit.

\subsection{Man in the middle}\label{MIM}

To find out anything about $\phi$, Eve needs to know the states that are used in Bob's measurements. Eve could attempt this by measuring the states in the quantum channel, but we saw above that this would be detected rapidly by Alice's checks on the test states. What if, instead, Eve jointly attacked both the quantum and classical channels so that when she measured a quantum state, she hid this by amending the classical data that Bob sends.

It is clear that it is never to Eve's advantage to swap the detector value 1 for 2 or 3 in the classical data as this will only aid Alice in detecting her. Suppose instead that Eve attempts to hide the states that she has measured by changing the classical data that Bob sends so that these are not identified as tests, i.e. switching appropriate values of 2 and 3 with 1 in Fig.~\ref{fig:photon_SQRS}. In this way, Alice would not check any of the states that Eve had measured and she could, in principle, go undetected while gaining information about $\phi$.

One way of mitigating this in our scheme is for Alice and Bob to agree that a certain fraction of the states will be used for tests. This fraction can be publicly declared. Alice can then perform a statistical analysis of the number of test and measurement states declared in Bob's message. If it deviates from what is expected, this will give evidence that Eve is intervening. 

This means that, while Eve can cover up the quantum states she measures by swapping the classical data to ensure these are not checked as test states, the number of times she can do this is limited by her not wanting to reveal herself through the biased distribution of $\{1, 2, 3\}$ in the classical data. Eve will only want to swap 2 and 3 for 1 so the distribution will become skewed. The question is whether Eve will be detected before she gains meaningful information about $\phi$.

\begin{figure}[ht]
  \centering
    \includegraphics[width=.45\textwidth]{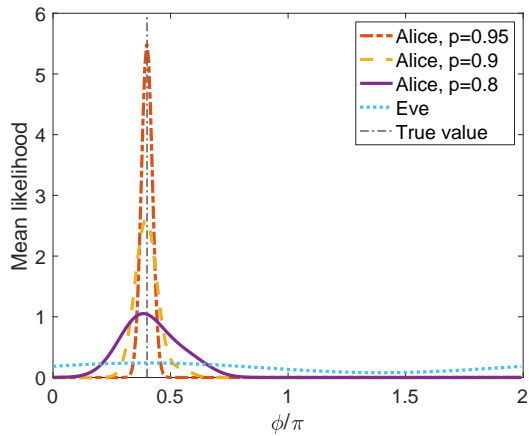}
    \caption{Bayesian prediction of $\phi$ for Alice and Eve  in the man in the middle attack scheme for a true value of $\phi$ of $0.4\pi$ and results have been averaged $10^{6}$ times. Eve has 1 information gaining measurement while Alice uses $p\in\{0.8,0.9,0.95\}$ which allows her to use 20, 90 and 280 information-gaining measurements respectively. If Eve uses a measure and resend attack on the quantum channel half of the time she measures in the wrong basis giving Alice the wrong result for the affected measurement.}
    \label{Fig:MITMalt}
\end{figure}

The $\{1, 2, 3\}$ data sent to Alice has a binomial distribution with variance $\sigma^2 = \mu p(1-p)$, where $\mu$ is the number of states and $p$ is the probability that a given state is used as a test, $\{2,3\}$. Eve's measurements will be detected if she swaps more than $~\sigma = \sqrt{\mu p(1-p)}$ elements of the data set, meaning she can measure $\sqrt{\mu (1-p)/p}$ states because only the fraction $p$ are test states and so need swapping. Overall, Eve will successfully gain information from $(1-p)\sqrt{\mu (1-p)/p}$ states because, of the states she detects, only the fraction $(1-p)$ will be used by Bob for measurements of $\phi$. We can limit the information Eve receives by setting the total number of information-gaining states she manages to measure and hide to 1, which then gives $\mu=p/(1-p)^3$. By comparison, Alice will have $n(1-p)$ information-gaining measurements. Taking the ratio of Eve's to Alice's information-gaining measurements, we get $(1-p)^2/p$. This is a monotonically decreasing function of $p$, so is minimised for $p=1$. However, this $p=1$ means that all the states are used for tests and Alice would get no information. Instead, for practical purposes we take $p$ to be close to 1 to both minimise the ratio and ensure that Alice can get information about $\phi$. As an example, we take $p=0.9$, which means that Alice gets 90 information-gaining states, while Eve gets only 1 (as  fixed above). Alice's and Eve's Bayesian predictions of $\phi$ are shown in Fig.~\ref{Fig:MITMalt} for $\phi = 0.4\pi$ and different values of $p$. We see that Alice correctly predicts this value with a clear peak, whereas Eve gains very little information. 

Eve could get around this problem by employing a measure and resend strategy and only altering the values 2 and 3 in the classical data that is sent to Alice. If Eve measures and resends quantum states that subsequently go down the fidelity checking path and are measured in a different basis to the one that Eve measured in, she could switch the values 2 and 3. This would hide what she had done because she can be sure that the subset of states Alice checks will not have been altered. This attack can be prevented by Alice using a secret basis as introduced in Section~\ref{authentication} making the scheme robust to man in the middle attacks of this sort.

\section{Conclusion}

We show a method of performing metrology at a remote site in a way that is secure from eavesdropping by using the indistinguishability of non-orthogonal quantum states to prevent Eve making measurements undetected on the quantum communication channel. We are able to ensure similar security and measurement capabilities as previous SQRS schemes without using entanglement. This provides advantages in practicality and bandwidth. We analyse the use of this protocol with limited data to highlight its practicality for real world applications where data may be limited by time, resource cost or security. We further highlight its efficiency by showing that by passing qubits through the sample multiple times we can get Heisenberg scaling of the measurement precision without using entanglement. We can correctly identify the correct likelihood peak in a $2\pi$ range by combining the results of a multipass test and a single pass test with sufficient data. 

We further highlight the efficiency of the method by showing that it may be applied to photonic coherent states. We show that our scheme is better protected against photon-splitting attacks than quantum cryptography schemes that must publicly announce their measurement basis like BB84. This could allow Alice and Bob to run SQRS using coherent states with a low mean photon number safe in the knowledge that even if there is a photon-splitting attack little useful information will be leaked. This information asymmetry may be further enhanced when Alice uses a well-chosen multipass-singlepass combination as Eve will not have enough data from the singlepass results to pick out the correct multipass peak and take advantage of the same Heisenberg scaling as Alice. 

We provide a detailed security analysis of the system, introducing features such as random path choice and path information delay to protect against spoofing and man in the middle type attacks and put limits on the effects of external influence in the classical communication channel. We also introduce secrets that are a part of the quantumness of the system and ensure the security of the scheme to man in the middle attacks where Eve tries to assumes the role of either Alice or Bob. Our results point to a new and practical way of implementing SQRS, which could be extended in the future to noisy scenarios\cite{Okane2020a}, multiple parameters and measurements on secure networks of multiple parties~\cite{Shettell2022a,Shettell2022b, Huang2019a}.

\section{Acknowledgements}

We acknowledge financial support for this research from DSTL under contract DSTLX1000146546.

\section{Author Declarations}
The authors have no conflicts to disclose.

\section{Data Availabilty Statement}
The data that support the findings of this study are available from the corresponding author upon reasonable request.

\appendix

\section{Circular statistics}\label{circStats}

Linear statistics consider a linear scale whereas directional statistics, in general, are performed by considering each data point as a vector in a higher dimensional space~\cite{DirectionalStats}. On a circle a data point can be represented by a vector, $\boldsymbol{x}$, parameterised by an angle, $\theta$
\begin{equation}
    \boldsymbol{x} = (\cos(\theta),\sin(\theta))^T,
\end{equation}
or equivalently by a complex number 
\begin{equation}
    z=e^{i\theta} = \cos(\theta) + i\sin(\theta). \label{eq:circz}
\end{equation}
The mean resultant length of a set of data $\{ z_n \}_{n=1}^{N}$, where each element is defined by Eq.~(\ref{eq:circz}), is
\begin{equation}
    \bar{\rho} = \frac{1}{N}\sum_{n=1}^{N} z_n.
\end{equation}
We calculate our likelihood functions, $\mathcal{L}(\theta)$, numerically using a grid approximation, by splitting the $2\pi$ support into $K\geq1000$ equally-sized bins such that $\theta_k = \theta_0+k/K,\ k\in\{0,1,2,...,K-1\}$ and calculating the value of the likelihood function for each bin. Thus, the mean resultant vector of our likelihood function is
\begin{equation}
    \bar{\rho} = \frac{1}{K}\sum_{k=1}^{K} \mathcal{L}(\theta_k)z_k,
\end{equation}
with mean direction 
\begin{equation}
    \bar{\theta} = \arg(\bar{\rho})
\end{equation}
and mean resultant length
\begin{equation}
    \bar{R} = |\bar{\rho}|.
\end{equation}
Like the linear mean, the mean direction can be used as an estimator for a parameter given a likelihood function. The mean resultant length $0\leq\bar{R}\leq 1$ has extrema representing the data being evenly spread around a circle and where all results are at the same data point respectively. Thus, it is a measure of the width of the likelihood function and can be used to calculate a circular analogue of the standard deviation
\begin{equation}
    \nu = \left( -2\log(\bar{R})\right)^{1/2},
\end{equation}
which is equivalent to the linear standard deviation for sufficiently narrow, $\bar{R}\sim 1$, distributions.



%
%

%


\bibliography{myBib.bib}

\providecommand{\noopsort}[1]{}\providecommand{\singleletter}[1]{#1}%
\begin{thebibliography}{36}%
\makeatletter
\providecommand \@ifxundefined [1]{%
 \@ifx{#1\undefined}
}%
\providecommand \@ifnum [1]{%
 \ifnum #1\expandafter \@firstoftwo
 \else \expandafter \@secondoftwo
 \fi
}%
\providecommand \@ifx [1]{%
 \ifx #1\expandafter \@firstoftwo
 \else \expandafter \@secondoftwo
 \fi
}%
\providecommand \natexlab [1]{#1}%
\providecommand \enquote  [1]{``#1''}%
\providecommand \bibnamefont  [1]{#1}%
\providecommand \bibfnamefont [1]{#1}%
\providecommand \citenamefont [1]{#1}%
\providecommand \href@noop [0]{\@secondoftwo}%
\providecommand \href [0]{\begingroup \@sanitize@url \@href}%
\providecommand \@href[1]{\@@startlink{#1}\@@href}%
\providecommand \@@href[1]{\endgroup#1\@@endlink}%
\providecommand \@sanitize@url [0]{\catcode `\\12\catcode `\$12\catcode
  `\&12\catcode `\#12\catcode `\^12\catcode `\_12\catcode `\%12\relax}%
\providecommand \@@startlink[1]{}%
\providecommand \@@endlink[0]{}%
\providecommand \url  [0]{\begingroup\@sanitize@url \@url }%
\providecommand \@url [1]{\endgroup\@href {#1}{\urlprefix }}%
\providecommand \urlprefix  [0]{URL }%
\providecommand \Eprint [0]{\href }%
\providecommand \doibase [0]{http://dx.doi.org/}%
\providecommand \selectlanguage [0]{\@gobble}%
\providecommand \bibinfo  [0]{\@secondoftwo}%
\providecommand \bibfield  [0]{\@secondoftwo}%
\providecommand \translation [1]{[#1]}%
\providecommand \BibitemOpen [0]{}%
\providecommand \bibitemStop [0]{}%
\providecommand \bibitemNoStop [0]{.\EOS\space}%
\providecommand \EOS [0]{\spacefactor3000\relax}%
\providecommand \BibitemShut  [1]{\csname bibitem#1\endcsname}%
\let\auto@bib@innerbib\@empty
\bibitem [{\citenamefont {Giovannetti}\ \emph {et~al.}(2011)\citenamefont
  {Giovannetti}, \citenamefont {Lloyd},\ and\ \citenamefont
  {Maccone}}]{Giovannetti2011-ma}%
  \BibitemOpen
  \bibfield  {author} {\bibinfo {author} {\bibfnamefont {V.}~\bibnamefont
  {Giovannetti}}, \bibinfo {author} {\bibfnamefont {S.}~\bibnamefont {Lloyd}},
  \ and\ \bibinfo {author} {\bibfnamefont {L.}~\bibnamefont {Maccone}},\ }\href
  {\doibase 10.1038/nphoton.2011.35} {\bibfield  {journal} {\bibinfo  {journal}
  {Nature Photonics}\ }\textbf {\bibinfo {volume} {5}},\ \bibinfo {pages} {222}
  (\bibinfo {year} {2011})}\BibitemShut {NoStop}%
\bibitem [{\citenamefont {Pirandola}\ \emph {et~al.}(2020)\citenamefont
  {Pirandola}, \citenamefont {Andersen}, \citenamefont {Banchi}, \citenamefont
  {Berta}, \citenamefont {Bunandar}, \citenamefont {Colbeck}, \citenamefont
  {Englund}, \citenamefont {Gehring}, \citenamefont {Lupo}, \citenamefont
  {Ottaviani}, \citenamefont {Pereira}, \citenamefont {Razavi}, \citenamefont
  {Shaari}, \citenamefont {Tomamichel}, \citenamefont {Usenko}, \citenamefont
  {Vallone}, \citenamefont {Villoresi},\ and\ \citenamefont
  {Wallden}}]{Pirandola:20}%
  \BibitemOpen
  \bibfield  {author} {\bibinfo {author} {\bibfnamefont {S.}~\bibnamefont
  {Pirandola}}, \bibinfo {author} {\bibfnamefont {U.~L.}\ \bibnamefont
  {Andersen}}, \bibinfo {author} {\bibfnamefont {L.}~\bibnamefont {Banchi}},
  \bibinfo {author} {\bibfnamefont {M.}~\bibnamefont {Berta}}, \bibinfo
  {author} {\bibfnamefont {D.}~\bibnamefont {Bunandar}}, \bibinfo {author}
  {\bibfnamefont {R.}~\bibnamefont {Colbeck}}, \bibinfo {author} {\bibfnamefont
  {D.}~\bibnamefont {Englund}}, \bibinfo {author} {\bibfnamefont
  {T.}~\bibnamefont {Gehring}}, \bibinfo {author} {\bibfnamefont
  {C.}~\bibnamefont {Lupo}}, \bibinfo {author} {\bibfnamefont {C.}~\bibnamefont
  {Ottaviani}}, \bibinfo {author} {\bibfnamefont {J.~L.}\ \bibnamefont
  {Pereira}}, \bibinfo {author} {\bibfnamefont {M.}~\bibnamefont {Razavi}},
  \bibinfo {author} {\bibfnamefont {J.~S.}\ \bibnamefont {Shaari}}, \bibinfo
  {author} {\bibfnamefont {M.}~\bibnamefont {Tomamichel}}, \bibinfo {author}
  {\bibfnamefont {V.~C.}\ \bibnamefont {Usenko}}, \bibinfo {author}
  {\bibfnamefont {G.}~\bibnamefont {Vallone}}, \bibinfo {author} {\bibfnamefont
  {P.}~\bibnamefont {Villoresi}}, \ and\ \bibinfo {author} {\bibfnamefont
  {P.}~\bibnamefont {Wallden}},\ }\href {\doibase 10.1364/AOP.361502}
  {\bibfield  {journal} {\bibinfo  {journal} {Adv. Opt. Photon.}\ }\textbf
  {\bibinfo {volume} {12}},\ \bibinfo {pages} {1012} (\bibinfo {year}
  {2020})}\BibitemShut {NoStop}%
\bibitem [{\citenamefont {Giovannetti}\ \emph {et~al.}(2002)\citenamefont
  {Giovannetti}, \citenamefont {Lloyd},\ and\ \citenamefont
  {Maccone}}]{Giovanetti2002a}%
  \BibitemOpen
  \bibfield  {author} {\bibinfo {author} {\bibfnamefont {V.}~\bibnamefont
  {Giovannetti}}, \bibinfo {author} {\bibfnamefont {S.}~\bibnamefont {Lloyd}},
  \ and\ \bibinfo {author} {\bibfnamefont {L.}~\bibnamefont {Maccone}},\ }\href
  {\doibase 10.1103/PhysRevA.65.022309} {\bibfield  {journal} {\bibinfo
  {journal} {Phys. Rev. A}\ }\textbf {\bibinfo {volume} {65}},\ \bibinfo
  {pages} {022309} (\bibinfo {year} {2002})}\BibitemShut {NoStop}%
\bibitem [{\citenamefont {Kasai}\ \emph {et~al.}(2022)\citenamefont {Kasai},
  \citenamefont {Takeuchi}, \citenamefont {Hakoshima}, \citenamefont
  {Matsuzaki},\ and\ \citenamefont {Tokura}}]{Kasai2022a}%
  \BibitemOpen
  \bibfield  {author} {\bibinfo {author} {\bibfnamefont {H.}~\bibnamefont
  {Kasai}}, \bibinfo {author} {\bibfnamefont {Y.}~\bibnamefont {Takeuchi}},
  \bibinfo {author} {\bibfnamefont {H.}~\bibnamefont {Hakoshima}}, \bibinfo
  {author} {\bibfnamefont {Y.}~\bibnamefont {Matsuzaki}}, \ and\ \bibinfo
  {author} {\bibfnamefont {Y.}~\bibnamefont {Tokura}},\ }\href {\doibase
  10.7566/JPSJ.91.074005} {\bibfield  {journal} {\bibinfo  {journal} {Journal
  of the Physical Society of Japan}\ }\textbf {\bibinfo {volume} {91}},\
  \bibinfo {pages} {074005} (\bibinfo {year} {2022})}\BibitemShut {NoStop}%
\bibitem [{\citenamefont {K{\'o}m{\'a}r}\ \emph {et~al.}(2014)\citenamefont
  {K{\'o}m{\'a}r}, \citenamefont {Kessler}, \citenamefont {Bishof},
  \citenamefont {Jiang}, \citenamefont {S{\o}rensen}, \citenamefont {Ye},\ and\
  \citenamefont {Lukin}}]{Komar2014a}%
  \BibitemOpen
  \bibfield  {author} {\bibinfo {author} {\bibfnamefont {P.}~\bibnamefont
  {K{\'o}m{\'a}r}}, \bibinfo {author} {\bibfnamefont {E.~M.}\ \bibnamefont
  {Kessler}}, \bibinfo {author} {\bibfnamefont {M.}~\bibnamefont {Bishof}},
  \bibinfo {author} {\bibfnamefont {L.}~\bibnamefont {Jiang}}, \bibinfo
  {author} {\bibfnamefont {A.~S.}\ \bibnamefont {S{\o}rensen}}, \bibinfo
  {author} {\bibfnamefont {J.}~\bibnamefont {Ye}}, \ and\ \bibinfo {author}
  {\bibfnamefont {M.~D.}\ \bibnamefont {Lukin}},\ }\href {\doibase
  10.1038/nphys3000} {\bibfield  {journal} {\bibinfo  {journal} {Nature
  Physics}\ }\textbf {\bibinfo {volume} {10}},\ \bibinfo {pages} {582}
  (\bibinfo {year} {2014})}\BibitemShut {NoStop}%
\bibitem [{\citenamefont {Takeuchi}\ \emph {et~al.}(2019)\citenamefont
  {Takeuchi}, \citenamefont {Matsuzaki}, \citenamefont {Miyanishi},
  \citenamefont {Sugiyama},\ and\ \citenamefont {Munro}}]{Takeuchi2019a}%
  \BibitemOpen
  \bibfield  {author} {\bibinfo {author} {\bibfnamefont {Y.}~\bibnamefont
  {Takeuchi}}, \bibinfo {author} {\bibfnamefont {Y.}~\bibnamefont {Matsuzaki}},
  \bibinfo {author} {\bibfnamefont {K.}~\bibnamefont {Miyanishi}}, \bibinfo
  {author} {\bibfnamefont {T.}~\bibnamefont {Sugiyama}}, \ and\ \bibinfo
  {author} {\bibfnamefont {W.~J.}\ \bibnamefont {Munro}},\ }\href {\doibase
  10.1103/PhysRevA.99.022325} {\bibfield  {journal} {\bibinfo  {journal} {Phys.
  Rev. A}\ }\textbf {\bibinfo {volume} {99}},\ \bibinfo {pages} {022325}
  (\bibinfo {year} {2019})}\BibitemShut {NoStop}%
\bibitem [{\citenamefont {Okane}\ \emph {et~al.}(2021)\citenamefont {Okane},
  \citenamefont {Hakoshima}, \citenamefont {Takeuchi}, \citenamefont {Seki},\
  and\ \citenamefont {Matsuzaki}}]{Okane2020a}%
  \BibitemOpen
  \bibfield  {author} {\bibinfo {author} {\bibfnamefont {H.}~\bibnamefont
  {Okane}}, \bibinfo {author} {\bibfnamefont {H.}~\bibnamefont {Hakoshima}},
  \bibinfo {author} {\bibfnamefont {Y.}~\bibnamefont {Takeuchi}}, \bibinfo
  {author} {\bibfnamefont {Y.}~\bibnamefont {Seki}}, \ and\ \bibinfo {author}
  {\bibfnamefont {Y.}~\bibnamefont {Matsuzaki}},\ }\href {\doibase
  10.1103/PhysRevA.104.062610} {\bibfield  {journal} {\bibinfo  {journal}
  {Phys. Rev. A}\ }\textbf {\bibinfo {volume} {104}},\ \bibinfo {pages}
  {062610} (\bibinfo {year} {2021})}\BibitemShut {NoStop}%
\bibitem [{\citenamefont {Huang}\ \emph {et~al.}(2019)\citenamefont {Huang},
  \citenamefont {Macchiavello},\ and\ \citenamefont {Maccone}}]{Huang2019a}%
  \BibitemOpen
  \bibfield  {author} {\bibinfo {author} {\bibfnamefont {Z.}~\bibnamefont
  {Huang}}, \bibinfo {author} {\bibfnamefont {C.}~\bibnamefont {Macchiavello}},
  \ and\ \bibinfo {author} {\bibfnamefont {L.}~\bibnamefont {Maccone}},\ }\href
  {\doibase 10.1103/PhysRevA.99.022314} {\bibfield  {journal} {\bibinfo
  {journal} {Phys. Rev. A}\ }\textbf {\bibinfo {volume} {99}},\ \bibinfo
  {pages} {022314} (\bibinfo {year} {2019})}\BibitemShut {NoStop}%
\bibitem [{\citenamefont {Xie}\ \emph {et~al.}(2018)\citenamefont {Xie},
  \citenamefont {Xu}, \citenamefont {Chen},\ and\ \citenamefont
  {Wang}}]{Xie2018a}%
  \BibitemOpen
  \bibfield  {author} {\bibinfo {author} {\bibfnamefont {D.}~\bibnamefont
  {Xie}}, \bibinfo {author} {\bibfnamefont {C.}~\bibnamefont {Xu}}, \bibinfo
  {author} {\bibfnamefont {J.}~\bibnamefont {Chen}}, \ and\ \bibinfo {author}
  {\bibfnamefont {A.~M.}\ \bibnamefont {Wang}},\ }\href {\doibase
  10.1007/s11128-018-1884-z} {\bibfield  {journal} {\bibinfo  {journal}
  {Quantum Information Processing}\ }\textbf {\bibinfo {volume} {17}},\
  \bibinfo {pages} {116} (\bibinfo {year} {2018})}\BibitemShut {NoStop}%
\bibitem [{\citenamefont {Yin}\ \emph {et~al.}(2020)\citenamefont {Yin},
  \citenamefont {Takeuchi}, \citenamefont {Zhang}, \citenamefont {Yin},
  \citenamefont {Matsuzaki}, \citenamefont {Peng}, \citenamefont {Xu},
  \citenamefont {Xu}, \citenamefont {Tang}, \citenamefont {Zhou}, \citenamefont
  {Chen}, \citenamefont {Li},\ and\ \citenamefont {Guo}}]{Yin2020a}%
  \BibitemOpen
  \bibfield  {author} {\bibinfo {author} {\bibfnamefont {P.}~\bibnamefont
  {Yin}}, \bibinfo {author} {\bibfnamefont {Y.}~\bibnamefont {Takeuchi}},
  \bibinfo {author} {\bibfnamefont {W.-H.}\ \bibnamefont {Zhang}}, \bibinfo
  {author} {\bibfnamefont {Z.-Q.}\ \bibnamefont {Yin}}, \bibinfo {author}
  {\bibfnamefont {Y.}~\bibnamefont {Matsuzaki}}, \bibinfo {author}
  {\bibfnamefont {X.-X.}\ \bibnamefont {Peng}}, \bibinfo {author}
  {\bibfnamefont {X.-Y.}\ \bibnamefont {Xu}}, \bibinfo {author} {\bibfnamefont
  {J.-S.}\ \bibnamefont {Xu}}, \bibinfo {author} {\bibfnamefont {J.-S.}\
  \bibnamefont {Tang}}, \bibinfo {author} {\bibfnamefont {Z.-Q.}\ \bibnamefont
  {Zhou}}, \bibinfo {author} {\bibfnamefont {G.}~\bibnamefont {Chen}}, \bibinfo
  {author} {\bibfnamefont {C.-F.}\ \bibnamefont {Li}}, \ and\ \bibinfo {author}
  {\bibfnamefont {G.-C.}\ \bibnamefont {Guo}},\ }\href {\doibase
  10.1103/PhysRevApplied.14.014065} {\bibfield  {journal} {\bibinfo  {journal}
  {Phys. Rev. Applied}\ }\textbf {\bibinfo {volume} {14}},\ \bibinfo {pages}
  {014065} (\bibinfo {year} {2020})}\BibitemShut {NoStop}%
\bibitem [{\citenamefont {Fisher}(1922)}]{Fisher}%
  \BibitemOpen
  \bibfield  {author} {\bibinfo {author} {\bibfnamefont {R.~A.}\ \bibnamefont
  {Fisher}},\ }\href@noop {} {\bibfield  {journal} {\bibinfo  {journal} {Phil.
  Trans. Roy. Soc. A}\ }\textbf {\bibinfo {volume} {222}},\ \bibinfo {pages}
  {309} (\bibinfo {year} {1922})}\BibitemShut {NoStop}%
\bibitem [{\citenamefont {Hyllus}\ \emph {et~al.}(2012)\citenamefont {Hyllus},
  \citenamefont {Laskowski}, \citenamefont {Krischek}, \citenamefont
  {Schwemmer}, \citenamefont {Wieczorek}, \citenamefont {Weinfurter},
  \citenamefont {Pezz\'e},\ and\ \citenamefont
  {Smerzi}}]{Fihser_multiparticle_entanglement}%
  \BibitemOpen
  \bibfield  {author} {\bibinfo {author} {\bibfnamefont {P.}~\bibnamefont
  {Hyllus}}, \bibinfo {author} {\bibfnamefont {W.}~\bibnamefont {Laskowski}},
  \bibinfo {author} {\bibfnamefont {R.}~\bibnamefont {Krischek}}, \bibinfo
  {author} {\bibfnamefont {C.}~\bibnamefont {Schwemmer}}, \bibinfo {author}
  {\bibfnamefont {W.}~\bibnamefont {Wieczorek}}, \bibinfo {author}
  {\bibfnamefont {H.}~\bibnamefont {Weinfurter}}, \bibinfo {author}
  {\bibfnamefont {L.}~\bibnamefont {Pezz\'e}}, \ and\ \bibinfo {author}
  {\bibfnamefont {A.}~\bibnamefont {Smerzi}},\ }\href {\doibase
  10.1103/PhysRevA.85.022321} {\bibfield  {journal} {\bibinfo  {journal} {Phys.
  Rev. A}\ }\textbf {\bibinfo {volume} {85}},\ \bibinfo {pages} {022321}
  (\bibinfo {year} {2012})}\BibitemShut {NoStop}%
\bibitem [{\citenamefont {Bennett}\ and\ \citenamefont
  {Brassard}(2014)}]{Bennett2014QuantumCP}%
  \BibitemOpen
  \bibfield  {author} {\bibinfo {author} {\bibfnamefont {C.~H.}\ \bibnamefont
  {Bennett}}\ and\ \bibinfo {author} {\bibfnamefont {G.}~\bibnamefont
  {Brassard}},\ }\href@noop {} {\bibfield  {journal} {\bibinfo  {journal}
  {Theor. Comput. Sci.}\ }\textbf {\bibinfo {volume} {560}},\ \bibinfo {pages}
  {7} (\bibinfo {year} {2014})}\BibitemShut {NoStop}%
\bibitem [{\citenamefont {Hwang}(2003)}]{Decoy_state_origin}%
  \BibitemOpen
  \bibfield  {author} {\bibinfo {author} {\bibfnamefont {W.-Y.}\ \bibnamefont
  {Hwang}},\ }\href {\doibase 10.1103/PhysRevLett.91.057901} {\bibfield
  {journal} {\bibinfo  {journal} {Phys. Rev. Lett.}\ }\textbf {\bibinfo
  {volume} {91}},\ \bibinfo {pages} {057901} (\bibinfo {year}
  {2003})}\BibitemShut {NoStop}%
\bibitem [{\citenamefont {Ma}\ \emph {et~al.}(2005)\citenamefont {Ma},
  \citenamefont {Qi}, \citenamefont {Zhao},\ and\ \citenamefont
  {Lo}}]{Decoy_state_practical}%
  \BibitemOpen
  \bibfield  {author} {\bibinfo {author} {\bibfnamefont {X.}~\bibnamefont
  {Ma}}, \bibinfo {author} {\bibfnamefont {B.}~\bibnamefont {Qi}}, \bibinfo
  {author} {\bibfnamefont {Y.}~\bibnamefont {Zhao}}, \ and\ \bibinfo {author}
  {\bibfnamefont {H.-K.}\ \bibnamefont {Lo}},\ }\href {\doibase
  10.1103/PhysRevA.72.012326} {\bibfield  {journal} {\bibinfo  {journal} {Phys.
  Rev. A}\ }\textbf {\bibinfo {volume} {72}},\ \bibinfo {pages} {012326}
  (\bibinfo {year} {2005})}\BibitemShut {NoStop}%
\bibitem [{\citenamefont {Zhao}\ \emph {et~al.}(2006)\citenamefont {Zhao},
  \citenamefont {Qi}, \citenamefont {Ma}, \citenamefont {Lo},\ and\
  \citenamefont {Qian}}]{Decoy_state_first_exp}%
  \BibitemOpen
  \bibfield  {author} {\bibinfo {author} {\bibfnamefont {Y.}~\bibnamefont
  {Zhao}}, \bibinfo {author} {\bibfnamefont {B.}~\bibnamefont {Qi}}, \bibinfo
  {author} {\bibfnamefont {X.}~\bibnamefont {Ma}}, \bibinfo {author}
  {\bibfnamefont {H.-K.}\ \bibnamefont {Lo}}, \ and\ \bibinfo {author}
  {\bibfnamefont {L.}~\bibnamefont {Qian}},\ }\href {\doibase
  10.1103/PhysRevLett.96.070502} {\bibfield  {journal} {\bibinfo  {journal}
  {Phys. Rev. Lett.}\ }\textbf {\bibinfo {volume} {96}},\ \bibinfo {pages}
  {070502} (\bibinfo {year} {2006})}\BibitemShut {NoStop}%
\bibitem [{\citenamefont {Zhang}\ \emph {et~al.}(2008)\citenamefont {Zhang},
  \citenamefont {Zou}, \citenamefont {Li}, \citenamefont {Jin},\ and\
  \citenamefont {Guo}}]{Zhang2008a}%
  \BibitemOpen
  \bibfield  {author} {\bibinfo {author} {\bibfnamefont {S.}~\bibnamefont
  {Zhang}}, \bibinfo {author} {\bibfnamefont {X.}~\bibnamefont {Zou}}, \bibinfo
  {author} {\bibfnamefont {K.}~\bibnamefont {Li}}, \bibinfo {author}
  {\bibfnamefont {C.}~\bibnamefont {Jin}}, \ and\ \bibinfo {author}
  {\bibfnamefont {G.}~\bibnamefont {Guo}},\ }\href {\doibase
  10.1103/PhysRevA.77.044302} {\bibfield  {journal} {\bibinfo  {journal} {Phys.
  Rev. A}\ }\textbf {\bibinfo {volume} {77}},\ \bibinfo {pages} {044302}
  (\bibinfo {year} {2008})}\BibitemShut {NoStop}%
\bibitem [{\citenamefont {Rubio}\ \emph {et~al.}(2018)\citenamefont {Rubio},
  \citenamefont {Knott},\ and\ \citenamefont {Dunningham}}]{Rubio_2018}%
  \BibitemOpen
  \bibfield  {author} {\bibinfo {author} {\bibfnamefont {J.}~\bibnamefont
  {Rubio}}, \bibinfo {author} {\bibfnamefont {P.}~\bibnamefont {Knott}}, \ and\
  \bibinfo {author} {\bibfnamefont {J.}~\bibnamefont {Dunningham}},\ }\href
  {\doibase 10.1088/2399-6528/aaa234} {\bibfield  {journal} {\bibinfo
  {journal} {Journal of Physics Communications}\ }\textbf {\bibinfo {volume}
  {2}},\ \bibinfo {pages} {015027} (\bibinfo {year} {2018})}\BibitemShut
  {NoStop}%
\bibitem [{\citenamefont {Rubio}\ and\ \citenamefont
  {Dunningham}(2019)}]{Rubio_2019}%
  \BibitemOpen
  \bibfield  {author} {\bibinfo {author} {\bibfnamefont {J.}~\bibnamefont
  {Rubio}}\ and\ \bibinfo {author} {\bibfnamefont {J.}~\bibnamefont
  {Dunningham}},\ }\href {\doibase 10.1088/1367-2630/ab098b} {\bibfield
  {journal} {\bibinfo  {journal} {New Journal of Physics}\ }\textbf {\bibinfo
  {volume} {21}},\ \bibinfo {pages} {043037} (\bibinfo {year}
  {2019})}\BibitemShut {NoStop}%
\bibitem [{\citenamefont {Rubio}\ and\ \citenamefont
  {Dunningham}(2020)}]{Rubio_2020}%
  \BibitemOpen
  \bibfield  {author} {\bibinfo {author} {\bibfnamefont {J.}~\bibnamefont
  {Rubio}}\ and\ \bibinfo {author} {\bibfnamefont {J.}~\bibnamefont
  {Dunningham}},\ }\href {\doibase 10.1103/PhysRevA.101.032114} {\bibfield
  {journal} {\bibinfo  {journal} {Phys. Rev. A}\ }\textbf {\bibinfo {volume}
  {101}},\ \bibinfo {pages} {032114} (\bibinfo {year} {2020})}\BibitemShut
  {NoStop}%
\bibitem [{\citenamefont {Mardia}\ and\ \citenamefont
  {Jupp}(1999)}]{DirectionalStats}%
  \BibitemOpen
  \bibfield  {author} {\bibinfo {author} {\bibfnamefont {K.~V.}\ \bibnamefont
  {Mardia}}\ and\ \bibinfo {author} {\bibfnamefont {P.~E.}\ \bibnamefont
  {Jupp}},\ }\href@noop {} {\emph {\bibinfo {title} {Directional
  Statistics}}},\ edited by\ \bibinfo {editor} {\bibfnamefont {K.~V.}\
  \bibnamefont {Mardia}}\ and\ \bibinfo {editor} {\bibfnamefont {P.~E.}\
  \bibnamefont {Jupp}},\ Wiley Series in Probability and Statistics\ (\bibinfo
  {publisher} {John Wiley \& Sons},\ \bibinfo {address} {Chichester, England},\
  \bibinfo {year} {1999})\BibitemShut {NoStop}%
\bibitem [{\citenamefont {Boto}\ \emph {et~al.}(2000)\citenamefont {Boto},
  \citenamefont {Kok}, \citenamefont {Abrams}, \citenamefont {Braunstein},
  \citenamefont {Williams},\ and\ \citenamefont {Dowling}}]{Boto2020a}%
  \BibitemOpen
  \bibfield  {author} {\bibinfo {author} {\bibfnamefont {A.~N.}\ \bibnamefont
  {Boto}}, \bibinfo {author} {\bibfnamefont {P.}~\bibnamefont {Kok}}, \bibinfo
  {author} {\bibfnamefont {D.~S.}\ \bibnamefont {Abrams}}, \bibinfo {author}
  {\bibfnamefont {S.~L.}\ \bibnamefont {Braunstein}}, \bibinfo {author}
  {\bibfnamefont {C.~P.}\ \bibnamefont {Williams}}, \ and\ \bibinfo {author}
  {\bibfnamefont {J.~P.}\ \bibnamefont {Dowling}},\ }\href {\doibase
  10.1103/PhysRevLett.85.2733} {\bibfield  {journal} {\bibinfo  {journal}
  {Phys. Rev. Lett.}\ }\textbf {\bibinfo {volume} {85}},\ \bibinfo {pages}
  {2733} (\bibinfo {year} {2000})}\BibitemShut {NoStop}%
\bibitem [{\citenamefont {Dowling}(2008)}]{Dowling2008a}%
  \BibitemOpen
  \bibfield  {author} {\bibinfo {author} {\bibfnamefont {J.~P.}\ \bibnamefont
  {Dowling}},\ }\href {\doibase 10.1080/00107510802091298} {\bibfield
  {journal} {\bibinfo  {journal} {Contemporary Physics}\ }\textbf {\bibinfo
  {volume} {49}},\ \bibinfo {pages} {125} (\bibinfo {year} {2008})}\BibitemShut
  {NoStop}%
\bibitem [{\citenamefont {Lee}\ \emph {et~al.}(2002)\citenamefont {Lee},
  \citenamefont {Kok},\ and\ \citenamefont {Dowling}}]{Lee2002a}%
  \BibitemOpen
  \bibfield  {author} {\bibinfo {author} {\bibfnamefont {H.}~\bibnamefont
  {Lee}}, \bibinfo {author} {\bibfnamefont {P.}~\bibnamefont {Kok}}, \ and\
  \bibinfo {author} {\bibfnamefont {J.~P.}\ \bibnamefont {Dowling}},\ }\href
  {\doibase 10.1080/0950034021000011536} {\bibfield  {journal} {\bibinfo
  {journal} {Journal of Modern Optics}\ }\textbf {\bibinfo {volume} {49}},\
  \bibinfo {pages} {2325} (\bibinfo {year} {2002})}\BibitemShut {NoStop}%
\bibitem [{\citenamefont {Mitchell}\ \emph {et~al.}(2004)\citenamefont
  {Mitchell}, \citenamefont {Lundeen},\ and\ \citenamefont
  {Steinberg}}]{Mitchell2004a}%
  \BibitemOpen
  \bibfield  {author} {\bibinfo {author} {\bibfnamefont {M.~W.}\ \bibnamefont
  {Mitchell}}, \bibinfo {author} {\bibfnamefont {J.~S.}\ \bibnamefont
  {Lundeen}}, \ and\ \bibinfo {author} {\bibfnamefont {A.~M.}\ \bibnamefont
  {Steinberg}},\ }\href {\doibase 10.1038/nature02493} {\bibfield  {journal}
  {\bibinfo  {journal} {Nature}\ }\textbf {\bibinfo {volume} {429}},\ \bibinfo
  {pages} {161} (\bibinfo {year} {2004})}\BibitemShut {NoStop}%
\bibitem [{\citenamefont {Matthews}\ \emph {et~al.}(2011)\citenamefont
  {Matthews}, \citenamefont {Politi}, \citenamefont {Bonneau},\ and\
  \citenamefont {O'Brien}}]{Matthews2011a}%
  \BibitemOpen
  \bibfield  {author} {\bibinfo {author} {\bibfnamefont {J.~C.~F.}\
  \bibnamefont {Matthews}}, \bibinfo {author} {\bibfnamefont {A.}~\bibnamefont
  {Politi}}, \bibinfo {author} {\bibfnamefont {D.}~\bibnamefont {Bonneau}}, \
  and\ \bibinfo {author} {\bibfnamefont {J.~L.}\ \bibnamefont {O'Brien}},\
  }\href {\doibase 10.1103/PhysRevLett.107.163602} {\bibfield  {journal}
  {\bibinfo  {journal} {Phys. Rev. Lett.}\ }\textbf {\bibinfo {volume} {107}},\
  \bibinfo {pages} {163602} (\bibinfo {year} {2011})}\BibitemShut {NoStop}%
\bibitem [{\citenamefont {Leibfried}\ \emph {et~al.}(2005)\citenamefont
  {Leibfried}, \citenamefont {Knill}, \citenamefont {Seidelin}, \citenamefont
  {Britton}, \citenamefont {Blakestad}, \citenamefont {Chiaverini},
  \citenamefont {Hume}, \citenamefont {Itano}, \citenamefont {Jost},
  \citenamefont {Langer}, \citenamefont {Ozeri}, \citenamefont {Reichle},\ and\
  \citenamefont {Wineland}}]{Leibfried2005a}%
  \BibitemOpen
  \bibfield  {author} {\bibinfo {author} {\bibfnamefont {D.}~\bibnamefont
  {Leibfried}}, \bibinfo {author} {\bibfnamefont {E.}~\bibnamefont {Knill}},
  \bibinfo {author} {\bibfnamefont {S.}~\bibnamefont {Seidelin}}, \bibinfo
  {author} {\bibfnamefont {J.}~\bibnamefont {Britton}}, \bibinfo {author}
  {\bibfnamefont {R.~B.}\ \bibnamefont {Blakestad}}, \bibinfo {author}
  {\bibfnamefont {J.}~\bibnamefont {Chiaverini}}, \bibinfo {author}
  {\bibfnamefont {D.~B.}\ \bibnamefont {Hume}}, \bibinfo {author}
  {\bibfnamefont {W.~M.}\ \bibnamefont {Itano}}, \bibinfo {author}
  {\bibfnamefont {J.~D.}\ \bibnamefont {Jost}}, \bibinfo {author}
  {\bibfnamefont {C.}~\bibnamefont {Langer}}, \bibinfo {author} {\bibfnamefont
  {R.}~\bibnamefont {Ozeri}}, \bibinfo {author} {\bibfnamefont
  {R.}~\bibnamefont {Reichle}}, \ and\ \bibinfo {author} {\bibfnamefont
  {D.~J.}\ \bibnamefont {Wineland}},\ }\href {\doibase 10.1038/nature04251}
  {\bibfield  {journal} {\bibinfo  {journal} {Nature}\ }\textbf {\bibinfo
  {volume} {438}},\ \bibinfo {pages} {639} (\bibinfo {year}
  {2005})}\BibitemShut {NoStop}%
\bibitem [{\citenamefont {Higgins}\ \emph {et~al.}(2007)\citenamefont
  {Higgins}, \citenamefont {Berry}, \citenamefont {Bartlett}, \citenamefont
  {Wiseman},\ and\ \citenamefont {Pryde}}]{Higgins2007a}%
  \BibitemOpen
  \bibfield  {author} {\bibinfo {author} {\bibfnamefont {B.~L.}\ \bibnamefont
  {Higgins}}, \bibinfo {author} {\bibfnamefont {D.~W.}\ \bibnamefont {Berry}},
  \bibinfo {author} {\bibfnamefont {S.~D.}\ \bibnamefont {Bartlett}}, \bibinfo
  {author} {\bibfnamefont {H.~M.}\ \bibnamefont {Wiseman}}, \ and\ \bibinfo
  {author} {\bibfnamefont {G.~J.}\ \bibnamefont {Pryde}},\ }\href {\doibase
  10.1038/nature06257} {\bibfield  {journal} {\bibinfo  {journal} {Nature}\
  }\textbf {\bibinfo {volume} {450}},\ \bibinfo {pages} {393} (\bibinfo {year}
  {2007})}\BibitemShut {NoStop}%
\bibitem [{\citenamefont {Berry}\ \emph {et~al.}(2009)\citenamefont {Berry},
  \citenamefont {Higgins}, \citenamefont {Bartlett}, \citenamefont {Mitchell},
  \citenamefont {Pryde},\ and\ \citenamefont {Wiseman}}]{Berry2009a}%
  \BibitemOpen
  \bibfield  {author} {\bibinfo {author} {\bibfnamefont {D.~W.}\ \bibnamefont
  {Berry}}, \bibinfo {author} {\bibfnamefont {B.~L.}\ \bibnamefont {Higgins}},
  \bibinfo {author} {\bibfnamefont {S.~D.}\ \bibnamefont {Bartlett}}, \bibinfo
  {author} {\bibfnamefont {M.~W.}\ \bibnamefont {Mitchell}}, \bibinfo {author}
  {\bibfnamefont {G.~J.}\ \bibnamefont {Pryde}}, \ and\ \bibinfo {author}
  {\bibfnamefont {H.~M.}\ \bibnamefont {Wiseman}},\ }\href {\doibase
  10.1103/PhysRevA.80.052114} {\bibfield  {journal} {\bibinfo  {journal} {Phys.
  Rev. A}\ }\textbf {\bibinfo {volume} {80}},\ \bibinfo {pages} {052114}
  (\bibinfo {year} {2009})}\BibitemShut {NoStop}%
\bibitem [{\citenamefont {Fei}\ \emph {et~al.}(2018)\citenamefont {Fei},
  \citenamefont {Meng}, \citenamefont {Gao}, \citenamefont {Ma},\ and\
  \citenamefont {Wang}}]{Fei2018a}%
  \BibitemOpen
  \bibfield  {author} {\bibinfo {author} {\bibfnamefont {Y.-Y.}\ \bibnamefont
  {Fei}}, \bibinfo {author} {\bibfnamefont {X.-D.}\ \bibnamefont {Meng}},
  \bibinfo {author} {\bibfnamefont {M.}~\bibnamefont {Gao}}, \bibinfo {author}
  {\bibfnamefont {Z.}~\bibnamefont {Ma}}, \ and\ \bibinfo {author}
  {\bibfnamefont {H.}~\bibnamefont {Wang}},\ }\href {\doibase
  10.1140/epjd/e2018-90110-3} {\bibfield  {journal} {\bibinfo  {journal} {The
  European Physical Journal D}\ }\textbf {\bibinfo {volume} {72}},\ \bibinfo
  {pages} {107} (\bibinfo {year} {2018})}\BibitemShut {NoStop}%
\bibitem [{\citenamefont {Huttner}\ \emph {et~al.}(1995)\citenamefont
  {Huttner}, \citenamefont {Imoto}, \citenamefont {Gisin},\ and\ \citenamefont
  {Mor}}]{Huttner1995a}%
  \BibitemOpen
  \bibfield  {author} {\bibinfo {author} {\bibfnamefont {B.}~\bibnamefont
  {Huttner}}, \bibinfo {author} {\bibfnamefont {N.}~\bibnamefont {Imoto}},
  \bibinfo {author} {\bibfnamefont {N.}~\bibnamefont {Gisin}}, \ and\ \bibinfo
  {author} {\bibfnamefont {T.}~\bibnamefont {Mor}},\ }\href {\doibase
  10.1103/PhysRevA.51.1863} {\bibfield  {journal} {\bibinfo  {journal} {Phys.
  Rev. A}\ }\textbf {\bibinfo {volume} {51}},\ \bibinfo {pages} {1863}
  (\bibinfo {year} {1995})}\BibitemShut {NoStop}%
\bibitem [{\citenamefont {Liu}\ \emph {et~al.}(2011)\citenamefont {Liu},
  \citenamefont {Sun}, \citenamefont {Liang},\ and\ \citenamefont
  {Yuan}}]{Liu2011a}%
  \BibitemOpen
  \bibfield  {author} {\bibinfo {author} {\bibfnamefont {W.-T.}\ \bibnamefont
  {Liu}}, \bibinfo {author} {\bibfnamefont {S.-H.}\ \bibnamefont {Sun}},
  \bibinfo {author} {\bibfnamefont {L.-M.}\ \bibnamefont {Liang}}, \ and\
  \bibinfo {author} {\bibfnamefont {J.-M.}\ \bibnamefont {Yuan}},\ }\href
  {\doibase 10.1103/PhysRevA.83.042326} {\bibfield  {journal} {\bibinfo
  {journal} {Phys. Rev. A}\ }\textbf {\bibinfo {volume} {83}},\ \bibinfo
  {pages} {042326} (\bibinfo {year} {2011})}\BibitemShut {NoStop}%
\bibitem [{\citenamefont {Aharonovich}\ \emph {et~al.}(2016)\citenamefont
  {Aharonovich}, \citenamefont {Englund},\ and\ \citenamefont
  {Toth}}]{Aharonovich2016}%
  \BibitemOpen
  \bibfield  {author} {\bibinfo {author} {\bibfnamefont {I.}~\bibnamefont
  {Aharonovich}}, \bibinfo {author} {\bibfnamefont {D.}~\bibnamefont
  {Englund}}, \ and\ \bibinfo {author} {\bibfnamefont {M.}~\bibnamefont
  {Toth}},\ }\href {\doibase 10.1038/nphoton.2016.186} {\bibfield  {journal}
  {\bibinfo  {journal} {Nature Photonics}\ }\textbf {\bibinfo {volume} {10}},\
  \bibinfo {pages} {631} (\bibinfo {year} {2016})}\BibitemShut {NoStop}%
\bibitem [{\citenamefont {Wang}\ \emph {et~al.}(2021)\citenamefont {Wang},
  \citenamefont {Zhang}, \citenamefont {Wang}, \citenamefont {Cheng},
  \citenamefont {Yang}, \citenamefont {Tang}, \citenamefont {Yan},
  \citenamefont {Tang}, \citenamefont {Liu}, \citenamefont {Yu}, \citenamefont
  {Zhang},\ and\ \citenamefont {Pan}}]{Wang2021}%
  \BibitemOpen
  \bibfield  {author} {\bibinfo {author} {\bibfnamefont {L.-J.}\ \bibnamefont
  {Wang}}, \bibinfo {author} {\bibfnamefont {K.-Y.}\ \bibnamefont {Zhang}},
  \bibinfo {author} {\bibfnamefont {J.-Y.}\ \bibnamefont {Wang}}, \bibinfo
  {author} {\bibfnamefont {J.}~\bibnamefont {Cheng}}, \bibinfo {author}
  {\bibfnamefont {Y.-H.}\ \bibnamefont {Yang}}, \bibinfo {author}
  {\bibfnamefont {S.-B.}\ \bibnamefont {Tang}}, \bibinfo {author}
  {\bibfnamefont {D.}~\bibnamefont {Yan}}, \bibinfo {author} {\bibfnamefont
  {Y.-L.}\ \bibnamefont {Tang}}, \bibinfo {author} {\bibfnamefont
  {Z.}~\bibnamefont {Liu}}, \bibinfo {author} {\bibfnamefont {Y.}~\bibnamefont
  {Yu}}, \bibinfo {author} {\bibfnamefont {Q.}~\bibnamefont {Zhang}}, \ and\
  \bibinfo {author} {\bibfnamefont {J.-W.}\ \bibnamefont {Pan}},\ }\href
  {\doibase 10.1038/s41534-021-00400-7} {\bibfield  {journal} {\bibinfo
  {journal} {npj Quantum Information}\ }\textbf {\bibinfo {volume} {7}},\
  \bibinfo {pages} {67} (\bibinfo {year} {2021})}\BibitemShut {NoStop}%
\bibitem [{\citenamefont {Shettell}\ \emph
  {et~al.}(2022{\natexlab{a}})\citenamefont {Shettell}, \citenamefont
  {Kashefi},\ and\ \citenamefont {Markham}}]{Shettell2022a}%
  \BibitemOpen
  \bibfield  {author} {\bibinfo {author} {\bibfnamefont {N.}~\bibnamefont
  {Shettell}}, \bibinfo {author} {\bibfnamefont {E.}~\bibnamefont {Kashefi}}, \
  and\ \bibinfo {author} {\bibfnamefont {D.}~\bibnamefont {Markham}},\ }\href
  {\doibase 10.1103/PhysRevA.105.L010401} {\bibfield  {journal} {\bibinfo
  {journal} {Phys. Rev. A}\ }\textbf {\bibinfo {volume} {105}},\ \bibinfo
  {pages} {L010401} (\bibinfo {year} {2022}{\natexlab{a}})}\BibitemShut
  {NoStop}%
\bibitem [{\citenamefont {Shettell}\ \emph
  {et~al.}(2022{\natexlab{b}})\citenamefont {Shettell}, \citenamefont
  {Hassani},\ and\ \citenamefont {Markham}}]{Shettell2022b}%
  \BibitemOpen
  \bibfield  {author} {\bibinfo {author} {\bibfnamefont {N.}~\bibnamefont
  {Shettell}}, \bibinfo {author} {\bibfnamefont {M.}~\bibnamefont {Hassani}}, \
  and\ \bibinfo {author} {\bibfnamefont {D.}~\bibnamefont {Markham}},\ }\href
  {https://arxiv.org/abs/2207.14450} {\bibfield  {journal} {\bibinfo  {journal}
  {preprint arXiv:2207.14450}\ } (\bibinfo {year}
  {2022}{\natexlab{b}})}\BibitemShut {NoStop}%
\end{thebibliography}%

\end{document}